\documentclass[useAMS,usenatbib,onecolumn]{mn2e}
\usepackage{aas_macros}
\usepackage{graphicx}
\title[Relativistic self-similar expansion II]{Relativistic Expansion of
Magnetic Loops at the Self-similar Stage II: \\
Magnetized outflows interacting with the ambient plasma}
\author[H. R. Takahashi, E. Asano, R. Matsumoto]{Hiroyuki
R. Takahashi$^{1}$\thanks{E-mail:takahashi@cfca.jp}, Eiji Asano$^2$ and
Ryoji Matsumoto$^{3}$\\
$^{1}$Center for Computational Astrophysics,
National Astronomical Observatory of Japan, 2-21-1, Osawa, Mitaka,
Tokyo, 181-8588, Japan \\
$^{2}$Kwasan and Hida Observatories, Kyoto University,
17 Ohmine-cho, Kita Kazan, Yamashina-ku, Kyoto 607-8471, Japan\\
$^{3}$Department of Physics, Graduate School of Science, Chiba University,
1-33 Yayoi-cho, Inage-ku, Chiba 263-8522, Japan}
\begin{document}

\date{Accepted 2011 February 10. Received 2011 February 09; in original
form 2010 June 22}

\pagerange{\pageref{firstpage}--\pageref{lastpage}} \pubyear{2011}

\maketitle

\label{firstpage}

\begin{abstract}
We obtained self-similar solutions of relativistically expanding
magnetic loops by assuming axisymmetry and a purely radial flow. The 
stellar rotation and the magnetic fields in the ambient plasma are
neglected. 
We include the Newtonian gravity of the central star. 
These solutions are extended from those in our previous work
\citep*{2009MNRAS.394..547T} by taking into account discontinuities such
as the contact discontinuity and the shock.
The global plasma flow consists of three regions, the
outflowing region, the post shocked region, and the ambient plasma. 
They are divided by two discontinuities.
The solutions are characterized by the radial velocity, which plays a
 role of the self-similar parameter in our solutions. 
The shock Lorentz factor gradually increases with radius.
It can be approximately represented by the power 
of radius with the power law index of $0.25$. 

We also carried out magnetohydrodynamic (MHD) simulations of the
evolution of magnetic loops to study the stability and the generality
of our analytical solutions. 
We used the analytical solutions as the initial condition and the inner
boundary conditions. We confirmed that our solutions are stable
over the simulation time and that numerical results nicely
recover the analytical solutions.
We then carried out numerical simulations to study the generality of our
solutions by changing the power law index $\delta$ of the ambient plasma
density $\rho_0 \propto r^{-\delta}$. We alter the power law index
$\delta$ from $\delta\simeq 3.5$ in the analytical solutions.
The analytical solutions are used as the initial conditions inside the
 shock in all simulations.
We observed that the shock Lorentz factor increases with time when the
power law index is larger than $3$, while it decreases with time when
the power law index is smaller than $3$. 
The shock Lorentz factor $\Gamma_s$ can be expressed as $\Gamma_s
 \propto t^{(\delta-3)/2}$ where $\delta$ is the power law index of the
 ambient plasma.
These results are consistent
 with the analytical studies by \cite{1979ApJ...233..831S}.
\end{abstract}

\begin{keywords}
 relativity - stars: magnetic fields - stars: neutron - stars: flare
\end{keywords}

\section{Introduction}\label{intro}
Soft Gamma-ray Repeaters (SGRs) and Anomalous X-ray Pulsars (AXPs) are
believed to be young neutron stars with strong magnetic fields ($\sim
10^{15}~\mathrm{G}$). They are
categorized as magnetars (see, e.g., \citealt{2006csxs.book..547W};
\citealt{2008A&ARv..15..225M}, for review). 
Their rotation period $P$ and its time derivative $\dot{P}$ are $P \sim
10~\mathrm{s}$ and $\dot{P}\sim 10^{-10}~\mathrm{s}~\mathrm{s}^{-1}$,
respectively.  
A magnetic field strength inferred by assuming the dipole emission from
$P$ and $\dot{P}$ is about $10^{15}~\mathrm{Gauss}$. 
Persistent X-ray emissions with the luminosity
of $L_X\sim 10^{34}-10^{36}~\mathrm{erg~s}^{-1}$ are observed in SGRs
and AXPs. 
SGRs are identified by the hard X-ray bursts. 
Extraordinary energetic outbursts called giant flares are observed in
three SGRs. The burst energy in the SGR 1806-20 giant flares
on December 27, 2004, reached $\sim 10^{47}~\mathrm{erg}$
\citep{2005Natur.434.1110T}. 
Since SGRs and AXPs are not accretion powered sources,
it is believed that their activity is driven by dissipation of
their magnetic fields.

Strong magnetic fields are created by the dynamo mechanism
during the core collapse of a supernova progenitor. At this stage, the
star becomes unstable against
the convective mode since the entropy gradient becomes negative,
($\mathrm{d}S/\mathrm{d} r<0$) due to the neutrino cooling
\citep{1987ApJ...318L..57B, 1996ApJ...473L.111K}. Thus the
infant neutron star can store a large amount of magnetic energies. 
The magnetic fields are also amplified after the birth of the magnetar. 
The interior of the star rotates differentially at its birth when the
equation of state is stiff. The internal magnetic
fields can be amplified up to $10^{17} \mathrm{G}$ due to the dynamo
mechanisms \citep{1992ApJ...392L...9D}. 
As the magnetic helicity is accumulated inside the star, the Lorentz 
force exerted by the twisted magnetic fields balances with the rigidity 
of the crusts.
When the critical twists are accumulated, the crustal rigidity can no
longer sustain the Lorentz force by the strong magnetic fields.
The magnetic helicity is then injected into the magnetosphere. 
The resulting crustal motion induces the electric fields and it results in
creating the potential difference between the
foot-points of the magnetic loops. The particles injected from the
interior of the star are accelerated along the magnetic field lines due
to the potential difference. The accelerated particles initiate the avalanches
of the pair creation \citep{2007ApJ...657..967B}. These particles
carry the electric current, which twists the global magnetic fields.
When the dynamical equilibrium is lost by the accumulated magnetic
twists, the magnetic loops expand by the magnetic pressure gradient force. 
Inside the magnetic loops, a current sheet similar to 
that of solar flares is formed. The magnetic reconnections
taking place in the current sheet are responsible for the magnetic energy
release and resulting flares \citep{2006MNRAS.367.1594L}. 
Recent observations which indicate the topological change of the global
magnetic fields before and after the giant flares support these models
\citep{2001ApJ...552..748W}.

Motivated by the magnetar flare model, \citet{2005KITP...CONF..HP}
performed 2-dimensional relativistic force-free simulations of magnetar
flares by injecting the magnetic twists into the magnetosphere. They
showed that the initially dipole magnetic fields are twisted by the
foot-point motion and the loop magnetic fields then expand due to the
magnetic pressure gradient force. 
\citet{2007Chiba...1..1} carried out 2-dimensional
relativistic force-free simulations of expanding magnetic loops and
showed that the Lorentz factor defined by the
drift velocity $\bmath v_d = c(\bmath E \times \bmath B)/B^2$ exceeds 10
\citep[see,][for the definition of the drift
velocity]{1997PhRvE..56.2181U}. These simulations indicate
that the magnetic loops expand self-similarly. 

Such self-similar solutions have been found in analytical studies. 
In the framework of the force-free dynamics, 
\citet{2003astro.ph.12347L} obtained self-similar solutions of the
spherically expanding magnetic shell. 
\citet{2005MNRAS.359..725P} found self-similar solutions of the
relativistic force-free field in two dimensions. 
\cite{2008MNRAS.391..268G} derived relativistic self-similar
force-free solutions and analyzed them in detail. 
In the framework of the relativistic magnetohydrodynamics (MHD),
\citet{2002PhFl...14..963L}
found self-similar solutions of the spherically expanding magnetic
shells. 
Recently, \cite{2009MNRAS.394..547T} obtained self-similar solutions
of magnetic loops (not shell) by extending the non-relativistic solutions
obtained by \cite{1982ApJ...261..351L}.
\cite{2010arXiv1001.4209G} obtained solutions by ignoring the
gravity from the central star.
These authors studied outflows of the the magnetized plasma lifted up from the
central star. However, they did not consider the interaction between the
outflow and the interstellar matter.
\cite{1984ApJ...281..392L} obtained non-relativistic self-similar
solutions of the expanding magnetic loops interacting with the
interstellar matter. In their models, 
the outflows and the ambient plasma are divided by a contact
discontinuity. The forward propagating wave forms another discontinuity
(shock). 
This solution is useful to understand the coronal mass ejections
in solar flares. 
\cite{1992ApJ...388..415S} employed this solution as a test problem 
to check the validity and accuracy of axisymmetric MHD codes.

In this paper, we extend the analytic solutions given by
\cite{2009MNRAS.394..547T} by including the contact
discontinuity and the shock by extending the non-relativistic model by
\cite{1984ApJ...281..392L} to the relativistic regime.

This paper is organized as follows. In \S~\ref{derivation}, we summarize
the basic equations of self-similar relativistic MHD equations given by
\cite{2009MNRAS.394..547T}. In \S~\ref{expsol}, we show the solutions
of these equations including the two discontinuities. These solutions
represent the relativistic coronal mass ejection from the central
star. Such solutions are expected to explain the giant flares in
magnetars. The physical properties of the solutions are shown in
\S~\ref{Property}. 
We also carried out the 2-dimensional relativistic
MHD simulations to study the stability of the solutions. 
The analytical solutions shown in \S~\ref{expsol} are used as the
initial and boundary conditions for simulations. These results are shown in
\S~\ref{numerical}. We summarize our results in \S~\ref{summary}.

\section[]{Basic equations of self-similar relativistic MHD}\label{derivation}
In the following, we take the light speed as unity. The complete set of
the relativistic ideal MHD equations is 
\begin{equation}
 \frac{\partial }{\partial t}(\gamma \rho)+\nabla\cdot(\gamma\rho \bmath
  v)=0, \label{eq:MHDcont}
\end{equation}
\begin{equation}
 \rho\gamma\left[\frac{\partial}{\partial t}+(\bmath v \cdot \nabla)
						\right]\left(h\gamma\bmath
 v\right)= -\nabla p
 +\rho_e \bmath E+\bmath j \times \bmath B-\frac{G M \rho h \gamma^2 }{r^2} \bmath e_r,\label{eq:MHDeom}
\end{equation}
\begin{equation}
 \left[\frac{\partial}{\partial t}+(\bmath v \cdot
 \nabla)\right]\left(\ln\frac{p}{\rho^\Gamma}\right)=0,
 \label{eq:MHDentropy}
\end{equation}
\begin{equation}
 \nabla \cdot \bmath E=4\pi \rho_e, \label{eq:Gauss}
\end{equation}
\begin{equation}
 \nabla \cdot \bmath B = 0, \label{eq:nomonopole}
\end{equation}
\begin{equation}
 \frac{\partial \bmath B}{\partial t}+\nabla \times \bmath E=0,\label{eq:Faraday}
\end{equation}
\begin{equation}
 \frac{\partial \bmath E}{\partial t}=\nabla \times \bmath B-4\pi\bmath j,\label{eq:Ampere}
\end{equation}
\begin{equation}
 \bmath E=-\bmath v\times \bmath B,\label{eq:mhdcond}
\end{equation}
where $\bmath E, \bmath B,\bmath j, \bmath v,\gamma,\rho_e, \rho,
p, \Gamma$ are the electric field, the magnetic field, the current density,
the velocity, the Lorentz factor, the charge density, the mass density,
the pressure and the specific heat ratio, respectively. The vector
$\bmath e_r$ is a unit vector in the radial direction. Newtonian gravity
of the central star is
included as an external force.
Here $G$ is the gravitational constant, and
$r$ is the distance from the centre of the star.
We replaced the gravitational force $-G M\rho \gamma/r^2$ in
\cite{2009MNRAS.394..547T} with $-GM\rho h\gamma^2/r^2$, which 
can treat the gravitational force of the relativistic plasma more
properly \citep{2010arXiv1001.4209G}.

The relativistic specific enthalpy including the rest mass energy $h$ is
written as
\begin{equation}
 h=\frac{\epsilon+p}{\rho}=1+\frac{\Gamma}{\Gamma-1}\frac{p}{\rho}
  \equiv 1 + h_N,
\end{equation}
where $\epsilon$ is the energy density of the matter including the
photon energy coupled with the plasma, and $h_N=\Gamma
p/[(\Gamma-1)\rho]$ is the non-relativistic specific thermal enthalpy.

In the following, we take $\Gamma=4/3$ which corresponds to the
relativistic radiation pressure dominant plasma. 
Thus we can study the evolution of a fireball confined by magnetic fields.

We ignore the stellar rotation and assume a purely radial flow. 
We also assume axisymmetry. The magnetic fields in spherical coordinates
($r, \theta, \phi$) are then expressed in
terms of two independent scalar functions $\tilde{A}$ and $B$ as 
\begin{equation}
 \bmath B=\frac{1}{r \sin\theta}\left(\frac{1}{r}\frac{\partial
				  \tilde{A}}{\partial \theta}, -\frac{\partial
				  \tilde{A}}{\partial r}, B\right),\label{def:A}
\end{equation}

In the following, we assume that the evolution of the magnetic loops
can be described by a Lagrangian coordinate $\eta$:
\begin{equation}
 \eta \equiv \frac{r}{Z(t)}, \label{eq:defeta} 
\end{equation}
where $Z(t)$ is a scale function of time. 
We further assume that the flux function $\tilde{A}$ evolves with time
$t$ and radial distance $r$ through the
Lagrangian coordinate $\eta$, as
\begin{equation}
 \tilde{A}(t, r, \theta) = \tilde{A}(\eta, \theta). \label{eq:assumeA}
\end{equation}

The MHD equations are then written as
\begin{equation}
 \frac{\mathrm{d}^2 Z(t)}{\mathrm{d} t^2} = 0, \label{eq:Zt}
\end{equation}
\begin{equation}
 v_r = \dot Z(t) \eta, \label{eq:vel}
\end{equation}
\begin{equation}
 p(t, r, \theta) = \frac{P(\eta, \theta)}{Z^4(t)}, \label{eq:p}
\end{equation}
\begin{equation}
 \rho(t, r, \theta) \gamma = \frac{D(\eta, \theta)}{Z^3(t)}, \label{eq:rho}
\end{equation}
\begin{equation}
 B(t, r,\theta) = \frac{Q(\eta, \theta)}{Z(t)}, \label{eq:B}
\end{equation}
\begin{equation}
\frac{4\gamma^2v_r^2P}{\eta}=
\frac{\partial P}{\partial \eta}
 +\frac{1}{4\pi \eta^2 \sin^2\theta}
 \left\{\frac{\partial\tilde{A}}{\partial \eta}
 \left[\hat{\mathcal{L}}_{(\eta,\theta)}\tilde{A}
  -\frac{\partial}{\partial \eta} 
  \left(\dot{Z}^2\eta^2 \frac{\partial\tilde{A}}{\partial \eta}\right)\right]
 +Q \frac{\partial}{\partial \eta}\left(\frac{Q}{\gamma^2}
				  \right)\right\}
+\frac{GMD\gamma}{\eta^2}\left(1+\frac{4\gamma P}{ZD}\right)\label{eq:eomr}
\end{equation}
\begin{equation}
\frac{1}{\gamma^2}\frac{\partial \tilde{A}}{\partial
  \eta}\frac{\partial Q}{\partial \theta}-\frac{\partial
  \tilde{A}}{\partial\theta}\frac{\partial}{\partial
  \eta}\left(\frac{Q}{\gamma^2}\right)=0,\label{eq:eomphi}
\end{equation}
\begin{equation}
 4\pi\eta^2\sin^2\theta\frac{\partial P}{\partial\theta}+\frac{\partial
  \tilde{A}}{\partial
  \theta}\left[\hat{\mathcal{L}}_{(\eta,\theta)}\tilde{A}
	  -\frac{\partial}{\partial
	  \eta}\left(\eta^2\frac{\partial \tilde{A}}{\partial \eta}\right)
	 \right]+\frac{Q}{\gamma^2}\frac{\partial
  Q}{\partial\theta}=0,\label{eq:eomtheta}
\end{equation}
where $\hat{\mathcal{L}}_{(\eta, \theta)}$  is an operator: 
\begin{equation}
 \hat{\mathcal{L}}_{(\eta, \theta)} \equiv \frac{\partial^2}{\partial
  \eta^2}+\frac{\sin\theta}{\eta^2}\frac{\partial }{\partial
  \theta}\left(\frac{1}{\sin\theta}\frac{\partial }{\partial \theta}\right),\label{def:L}
\end{equation}
\citep[see,][for derivation]{2009MNRAS.394..547T}.
We can readily solve equation (\ref{eq:Zt}) as 
\begin{equation}
 Z(t)=\sqrt{\xi} t, \label{sol:Z}
\end{equation}
where $\xi$ is a constant of integral. 
The radial velocity is obtained from equation (\ref{eq:vel}) as
\begin{equation}
 v_r=\sqrt{\xi } \eta. \label{sol:v}
\end{equation}
By substituting equation (\ref{sol:Z}) into equation
(\ref{eq:eomphi}), we obtain the function $Q$ which has a form:
\begin{equation}
 Q(\eta, \theta) = \frac{g(\tilde{A})}{1-\eta^2},\label{def:g}
\end{equation}
where $g(\tilde{A})$ is an arbitrary function of $\tilde{A}$.

We note that the last term in equation (\ref{eq:eomr}) includes
$Z(t)=\sqrt{\xi}t$, which is a function of time. Thus this
term does not admit the existence of the self-similar solutions because
the thermal enthalpy in the gravity term explicitly depends on
time. This violation of the self-similarity comes from the difference of
the scaling law between the density and the pressure (equations
\ref{eq:p} and \ref{eq:rho}, also, see \citealt{2010arXiv1001.4209G}).  
However, when the contribution of the thermal enthalpy on the gravity is
sufficiently small, we can obtain self-similar solutions of the
relativistic expansion. The ratio of the gravity for the thermal enthalpy
to the plasma inertia is written as
\begin{equation}
 \frac{{\rm gravity~for~the~thermal~enthalpy}}
  {{\rm plasma~inertia}}
  =\frac{| GM \rho h_N\gamma^2 /r^2|}{|\rho \gamma D(\gamma hv)/Dt|}
  =\frac{r_g}{rv^2},\label{eq:fratio}
\end{equation}
where $r_g=GM$ is the gravitational radius. Thus when $r\gg r_g$, the
gravitational force for the thermal enthalpy $h_N$ is negligible when
$v\sim 1$. In this paper, we consider the evolution of the
relativistically expanding plasma at large distance where the gravity for
the thermal enthalpy $-GM\rho h_N\gamma^2/r^2$ can be neglected. 
In such region, the plasma inertia is sustained by the pressure gradient
and Lorentz forces.
When $r \ll R_s $, we assume that the rest mass energy density much
exceeds the thermal energy density (i.e., $h_N \ll 1$). 
In this regime, the gravity is expressed as $-GM\rho \gamma^2/r^2$. 
The Lagrangian coordinate $\eta$ (see equation \ref{eq:defeta}) then
behaves as the self-similar parameter.

Without loss of generality, we can change independent variables of $P$
from $(\eta, \theta)$ to $(\eta,
\tilde{A})$ as
\begin{equation}
 P(\eta, \theta)=P(\eta, \tilde{A}). \label{eq:Pchange}
\end{equation}
By neglecting the thermal enthalpy in the gravitational force $-GM \rho
h \gamma^2/r^2$, equations (\ref{eq:eomtheta}) and (\ref{eq:eomr}) reduce
to the self-similar equations as

\begin{equation}
 \frac{\partial P}{\partial \tilde A}=
  -\frac{1}{4\pi \eta^2 \sin^2\theta}
  \left[\hat{\mathcal{L}}_{(\eta, \theta)}\tilde{A}
   -\frac{\partial}{\partial \eta}
   \left(\xi \eta^2  \frac{\partial \tilde A}{\partial\eta}\right)
   +\frac{g(\tilde{A})}{1-\xi \eta^2}\frac{\mathrm d
   g(\tilde{A})}{\mathrm d \tilde{A}}\right],
  \label{eq:eomtheta2}
\end{equation}
\begin{equation}
 D=\frac{\eta^2\sqrt{1-\xi\eta^2}}{GM}\left(\frac{4 \xi \eta P}{1-\xi
		     \eta^2}-\left.\frac{\partial P}{\partial\eta}
			     \right|_{\tilde{A}}\right). \label{eq:eomr2}
\end{equation}
Equations (\ref{def:g}), (\ref{eq:eomtheta2}) and (\ref{eq:eomr2}) are
the set of the self-similar MHD equations. 
The explicit solutions can be constructed as follows. First
we prescribe an arbitrary function $\tilde{A}(\eta,\theta)$ and the
function $g(\tilde{A})$.
The pressure function $P$ is determined from equation
(\ref{eq:eomtheta2}) and the density function $D$ is obtained
from equation (\ref{eq:eomr2}). 

Before presenting the solutions of the self-similar MHD equations, 
we have to note that the self-similar
relativistic ideal MHD equations describe the free expansion of the
magnetized
plasma. By taking time derivative of $\bmath v$, we obtain
\begin{equation}
 \frac{D \bmath v}{D t}=0,
\end{equation}
where we used equations (\ref{eq:defeta}), (\ref{sol:Z}) and (\ref{sol:v}). 
The plasma is neither accelerated nor decelerated, but it expands with
the inertial speed keeping the force balance \citep[this can be
confirmed by inserting equation (\ref{sol:v}) into equation
(\ref{eq:MHDeom}), or, see equation (36) in][]{2009MNRAS.394..547T}.

\section{Relativistic Coronal Mass Ejection}{\label{expsol}}
In the previous section, we showed the set of the self-similar
relativistic MHD equations, which describes the plasma expanding with
the inertial speed. In this section, we obtain solutions of these equations by
imposing appropriate boundary conditions. 

\begin{figure}
 \center
 \includegraphics[width=8cm]{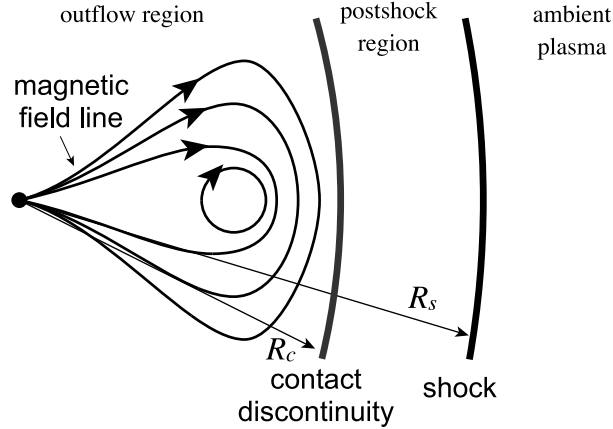}
 \caption{Schematic picture of our models of the relativistic coronal mass
 ejection. Ahead of the global magnetic loops, two discontinuities
 exist at $r=R_c(t)$ (contact discontinuity) and $r=R_s(t)$ (shock).}
 \label{fig:model}
\end{figure}

We adopt the simple model of magnetic
explosions according to Low models \citep[][see,
Fig.~{\ref{fig:model}}]{1984ApJ...281..392L}.
Equations (\ref{eq:eomtheta2}) and (\ref{eq:eomr2}) describe 
magnetized plasma outflowing from the central star (outflow region).
The outflow sweeps up an ambient plasma while it expands.
A contact surface would be formed which separates the outflowing
plasma with the swept-up ambient plasma. The contact surface is situated at
$r=R_c(t)$. Ahead of the contact
surface, a forward wave propagating into an undisturbed ambient plasma
forms a shock at $r=R_s(t)$. The swept-up ambient
plasma is accumulated in the region in $R_c(t) \lid r \lid R_s(t)$
(postshock region).

First, we consider the outflow region where the magnetized plasma is
lifted up from the central star. 
Various models of the magnetic field configurations have been proposed
\citep{1982ApJ...261..351L, 1984ApJ...281..392L, 2009MNRAS.394..547T}.
In this paper, we utilize the following flux function
\citep{1984ApJ...281..392L}
\begin{equation}
 \tilde{A}= A_0 \left[H_0+\sqrt{\frac{2}{\pi}}\left(\frac{\sin(\lambda
					      \eta)}{\lambda
					      \eta}-\cos(\lambda
					      \eta)\right)\right]\sin^2\theta\equiv
 A_0 f(\lambda \eta)\sin^2\theta,\label{sol:A}
\end{equation}
where
\begin{equation}
 f(x)=H_0+\sqrt{\frac{2}{\pi}}\left(\frac{\sin x}{x}-\cos x\right). \label{def:f}
\end{equation}
Here $A_0$, $H_0$, and $\lambda$ are constants.
The flux function given by equation (\ref{sol:A}) has local maxima (see
Fig.~4 in \citealt{1984ApJ...281..392L}). The maximum corresponds to the
centre of the flux ropes. 
This function can thus describe the coronal mass ejection from the
central star. Such relativistic coronal mass ejections can be applied to
giant flares in magnetars \citep{2006MNRAS.367.1594L}. 
We express the toroidal magnetic field $g$ as a power series of
$\tilde{A}$:
\begin{equation}
 g(\tilde{A})=\sum_n \alpha_n \tilde{A}^n, \label{sol:g}
\end{equation}
where $\alpha_n$ are constants. 
Note that the dependence of the toroidal magnetic
field on the polar angle is represented by $\sum_n \sin^{2n-1}\theta$ from equations
(\ref{def:A}), (\ref{eq:B}), (\ref{sol:A}) and (\ref{sol:g}).
This indicates that the power law index $n$ and $\alpha_n$ can be
interpreted as the Fourier modes of the toroidal
magnetic fields in polar direction and their amplitudes, respectively.
The modes and the amplitudes are determined by the boundary condition at
which the magnetic shear is injected into the magnetosphere. 
The shear injection begins before the self-similar expansion starts. 
In this paper, we do not consider the details of the shear injection
from the central star and leave them as free parameters since we
consider the self-similar stage.

The explicit forms of the magnetic fields are then written as
\begin{equation}
 B_r(t, r, \theta)=
  \frac{2A_0}{r^2}
  \left\{H_0+\sqrt{\frac{2}{\pi}}
   \left[\frac{\sin(\lambda\eta)}{\lambda\eta}-\cos(\lambda\eta)\right]\right\}
  \cos\theta,\label{sol:Br}
\end{equation}
\begin{equation}
 B_\theta(t, r,
  \theta)=-\frac{A_0}{r^2}\sqrt{\frac{2}{\pi}}
  \left[\cos(\lambda \eta)-\frac{1-(\lambda \eta)^2}{\lambda
   \eta}\sin(\lambda \eta)\right]\sin\theta,\label{sol:Btheta}
\end{equation}
\begin{equation}
 B_\phi(t, r, \theta)=\frac{1}{r^2\sin\theta}\frac{\eta}{1-\xi\eta^2}\sum_n
  \alpha_n \tilde{A}^n.\label{sol:Bphi}
\end{equation}

 The pressure function of the outflowing plasma $P_o$
is obtained by inserting equations (\ref{sol:A}) and (\ref{sol:g}) into
equation (\ref{eq:eomtheta2}) as
\begin{equation}
 P_o(\eta, \theta)=P_A(\eta, \theta) + P_Q (\eta, \theta) + P_i
  (\eta),\label{sol:Psum}
\end{equation}
where 
\begin{equation}
 P_A(\eta, \theta)=\frac{A_0 \tilde{A}(\eta, \theta)}{4\pi \eta^4}
  \left\{2 H_0+\sqrt{\frac{2}{\pi}}(\lambda \eta)
   \left[(1+\xi\eta^2)\sin(\lambda \eta)
    -(\lambda \eta)(1-\xi\eta^2)\cos(\lambda \eta)\right]\right\},
  \label{sol:PA}
\end{equation}
\begin{eqnarray}
 P_Q(\eta, \theta) =-\frac{1}{4\pi \eta^2 (1-\xi
  \eta^2)}\left\{\sum_{m+n\ne 1}\frac{m\alpha_m\alpha_n
	   A_0^{m+n}}{m+n-1}f^{m+n}(\lambda\eta)\sin^{2(m+n-1)}\theta
	   +\sum_{m+n=1}m\alpha_m\alpha_nA_0
	   f(\lambda \eta) \ln \tilde{A}(\eta,\theta)
\right\}. \label{sol:PQ}
\end{eqnarray}
Here $P_i$ arises from the integration. By substituting equations
(\ref{sol:Psum})-(\ref{sol:PQ}) into
equation (\ref{eq:eomr2}), the corresponding density function $D_o$ is
expressed as
\begin{equation}
 D_o(\eta, \theta)=D_A(\eta, \theta)+D_Q(\eta, \theta)+D_i(\eta),
  \label{sol:Dsum}
\end{equation}
\begin{equation}
 D_A(\eta, \theta)=\frac{A_0 \tilde{A}(\eta,\theta)}
  {4\pi GM \eta^3 \sqrt{1-\xi \eta^2}}\Xi(\eta),\label{DA}
\end{equation}
\begin{equation}  
 D_Q(\eta, \theta)=-\frac{(3-\xi\eta^2)f(\lambda \eta)
  -(1-\xi\eta^2)\left[H_0+\sqrt{\frac{2}{\pi}}(\lambda \eta )\sin(\lambda
		 \eta)\right]}
		 {4\pi GM\eta (1-\xi\eta^2)^\frac{3}{2}}\left\{\begin{array}{ll}
					      {\displaystyle
					       \sum_{m+n\ne
					       1}\frac{m\alpha_m\alpha_n
					       A_0\tilde A^{m+n-1}(\eta,
					       \theta)}{m+n-1}},\\
						      \hspace{4cm} {\rm
						       for \ \ } m+n\ne
						       1,\\
						      {\displaystyle \sum_{m+n=1}m\alpha_m\alpha_nA_0
						       \ln \tilde{A}(\eta,\theta)}, \\
						      \hspace{4cm} {\rm
						       for\ \ } m+n=1,
						       \end{array}\right.\label{sol:DQ}
\end{equation}
where the function $\Xi(\eta)$ is defined as
\begin{equation}
 \Xi(\eta)=8H_0+\sqrt{\frac{2}{\pi}}\lambda \eta\left\{
\left[(3-(\lambda\eta)^2)(1+(\xi \eta^2)^2)+2 \xi
	 \eta^2(1+(\lambda \eta)^2)\right]\sin(\lambda \eta)
-\lambda \eta(1-\xi \eta^2)(3+5\xi
  \eta^2)\cos(\lambda \eta)\right\}.\label{def:Xi}
\end{equation}
The function $D_i(\eta)$ describes the isotropic distribution of the plasma
which is related with $P_i(\eta)$ through equation (\ref{eq:eomr2}).

To determine functions $P_i$ and $D_i$, we need one more
relation between them.
We assume that there is no energy/mass injection from
the central star at this stage. 
Since the contact surface separates the outflow region from the
postshock region, the outflow plasma expands adiabatically.
We then obtain another relation between $P_i(\eta)$ and
$D_i(\eta)$ from the entropy conservation equation as
\begin{equation}
 \frac{p_i}{\rho_i^\frac{4}{3}}=\mathrm{const}=\frac{\nu}{4},
  \label{eq:adiabatic}
\end{equation}
where $p_i=Z(t)^4P_i$, $\rho_i=Z(t)^3 D_i$ and  $\nu$ is a constant.
Substituting equations (\ref{eq:p}), (\ref{eq:rho}), and (\ref{sol:v})
into equation (\ref{eq:adiabatic}), we obtain
\begin{equation}
 P_i=\frac{\nu}{4}\left(1-\xi\eta^2\right)^\frac{2}{3} D_i^{\frac{4}{3}},\label{eq:adiabatic2}
\end{equation}
Substituting equation (\ref{eq:adiabatic2}) into (\ref{eq:eomr2}), we
obtain the solutions,
\begin{equation}
 P_i(\eta)=\frac{1}{4\nu^3}
  \left(\frac{GM}{\eta}-\frac{\mu}{\sqrt{1-\xi\eta^2}}\right)^4,\label{sol:Pi}
\end{equation}
\begin{equation}
 D_i(\eta)=\frac{1}{\nu^3}\frac{1}{\sqrt{1-\xi\eta^2}}
  \left(\frac{GM}{\eta}-\frac{\mu}{\sqrt{1-\xi\eta^2}}\right)^3,\label{sol:Di}
\end{equation}
where $\mu$ is a constant of integral. These solutions describe the
isotropic outflowing plasma in the outflow region.

Next, we consider the post shock region ($R_c \lid r \lid R_s$). The shocked
plasma also moves in radial direction. We assume that the shocked plasma
evolves self-similarly and obeys the same set of
self-similar equations in the outflow region. Then the contact surface moves
in radial direction with a constant speed and the radius of the contact
surface $R_c$ is expressed as
\begin{equation}
 R_c(t)=\eta_c \sqrt{\xi} t, \label{def:eta_c}
\end{equation}
where $\eta_c$ is a constant.
We assumed that the magnetic fields in the ambient plasma can be
neglected.
From these assumptions, the shocked plasma obeys equation
(\ref{eq:eomr2}). We need another relation between the shocked
gas pressure $P_s$ and the shocked gas density $D_s$. Note that we
cannot use the adiabatic relation because the ambient plasma flows into the
postshock region from the shock surface at $r=R_s$. The forward shock
compresses and heats up the plasma, resulting in an increase in the entropy.
Thus the entropy in the shocked plasma should be determined by the shock
 condition at $r=R_s(t)$. 
Rather than evaluating the entropy variation by the shock, we consider
the jump conditions of the plasma density, the pressure, and the
velocity. The entropy variation is determined after imposing the
Rankin-Hugoniot relations between the undisturbed and shocked plasma
\citep{1984ApJ...281..381L}. 

By assuming the strong shocks, the relativistic Rankin-Hugoniot
relations are written as
\begin{equation}
 \left. p\right|_{r=R_s}=\left. \frac{2}{3}\Gamma_s^2\rho_0\right|_{r=R_s},
  \label{eq:RHp}
\end{equation}
\begin{equation}
 \left. \rho\gamma\right|_{r=R_s}=\left. 2\Gamma_s^2\rho_0 \right|_{r=R_s},
  \label{eq:RHr}
\end{equation}
\begin{equation}
 \left. \gamma^2 \right|_{r=R_s}= \frac{1}{2}\Gamma_s^2,
  \label{eq:RHg}
\end{equation}
with an accuracy of $\mathcal{O}\left(1/\gamma^2\right)$
\citep{1976PhFl...19.1130B, 1984ApJ...283..710K}.
$\Gamma_s$ and $\rho_0$ are the shock Lorentz factor and the
plasma mass density of the undisturbed ambient plasma,
respectively. Here we
ignore the thermal pressure in the undisturbed plasma
by assuming the strong shock. 

From equations (\ref{sol:v}) and (\ref{eq:RHg}), we obtain the time
evolution of the shock radius as
\begin{equation}
 R_s(t)=\frac{3R_0x_s}{\left(x_s+\sqrt{1+x_s^2}\right)^{\sqrt{2}}
  \left(\sqrt{2}x_s-\sqrt{1+x_s^2}\right)^2}, \label{sol:Rs}
\end{equation}
where $x_s\equiv R_s/t$ and $R_0$ is a constant of integral.
Combining equations (\ref{eq:RHp}) and (\ref{eq:RHr}), we obtain the
relation between the $P_s$ and $D_s$:
\begin{equation}
 P_s(\eta)=
  \frac{\sqrt{\xi}R_0}
{\left(\sqrt{\xi}\eta+\sqrt{1+\xi \eta^2}\right)^{\sqrt{2}}
  \left(\sqrt{2\xi}\eta-\sqrt{1+\xi\eta^2}\right)^2}
  D_s(\eta).\label{eq:PDs}
\end{equation}
Here we use equations (\ref{eq:p}), (\ref{eq:rho}), (\ref{sol:Rs}). 
Substituting equation (\ref{eq:PDs}) into equation (\ref{eq:eomr2}), 
$P_s$ and $D_s$ are expressed as
\begin{equation}
 P_s=\frac{P_0}{\left(1-\xi\eta^2\right)^2}
  K\left(\sqrt{\xi}\eta\right), \label{sol:Ps}
\end{equation}
\begin{equation}
 D_s=\frac{P_0}{\sqrt{\xi}R_0}
  \frac{\left(\sqrt{\xi}\eta+\sqrt{1+\xi \eta^2}\right)^{\sqrt{2}}}
  {\left(\sqrt{2\xi}\eta+\sqrt{1+\xi \eta^2}\right)^2}
  K\left(\sqrt{\xi}\eta\right), \label{sol:Ds}
\end{equation}
where $P_0$ is a constant of integral. The function $K$ is expressed
as
\begin{equation}
 K\left(\sqrt{\xi}\eta\right)\equiv\exp\left[-\frac{GM}{R_0}\int^{\sqrt{\xi}\eta}_{\sqrt{\xi}\eta_c}dx\frac{\left(x+\sqrt{1+x^2}\right)(1-x^2)^\frac{3}{2}}{x^2\left(\sqrt{2}x+\sqrt{1+x^2}\right)}\right].
 \label{def:K}
\end{equation}

The ambient plasma density is obtained by substituting equations
(\ref{sol:Rs}) and (\ref{sol:Ds}) into (\ref{eq:RHr}) as
\begin{equation}
 \rho_0(r)=\frac{3 P_0}{4 r^4} \frac{\eta_0^4}{(1-\xi\eta_0^2)}K\left(\sqrt{\xi}\eta_0\right).\label{sol:rho0}
\end{equation}
Here $\eta_0$ should be determined from the following equation,
\begin{equation}
 r=\frac{3R_0\sqrt{\xi}\eta_0}{\left(\sqrt{\xi}\eta_0+\sqrt{1+\xi \eta_0^2}\right)^{\sqrt{2}}
  \left(\sqrt{2\xi}\eta_0-\sqrt{1+\xi\eta_0^2}\right)^2}, \label{sol:rho0r}
\end{equation}
Note that the radial profile of the ambient plasma density is not
arbitrary but determined by equation (\ref{sol:rho0}).
Some authors derive the self-similar solutions by prescribing the
density profile of the ambient plasma as $\rho_0 \propto r^{-\delta}$,
where $\delta$ is a constant \citep{1976PhFl...19.1130B,
2006PhFl...18b7106S}.
In our approach, we first prescribe the self-similar variables given in
equation (\ref{eq:defeta}). Then the outflow velocity is obtained by
equation (\ref{eq:vel}). The ambient plasma density is determined by
applying the Rankin-Hugoniot relations at the shock. Thus the ambient plasma density cannot
have an arbitrary form, but it is uniquely determined.

Finally, we apply the boundary conditions at $r=R_c$. 
Since we assumed unmagnetized ambient plasma, the
magnetic fields should vanish at $r=R_c$. This condition can determine
the parameters $H_0$ and $\lambda$ in equation (\ref{sol:A}). 
The conditions that $B_r(r=R_c)=0$ and $B_\phi(r=R_c)=0$ are expressed as
\begin{equation}
 H_0=-\sqrt{\frac{2}{\pi}}
  \left(\frac{\sin(\lambda \eta_c)}
   {\lambda\eta_c}-\cos(\lambda\eta_c)\right),\label{eq:H0relation}
\end{equation}
from equations (\ref{sol:Br}) and (\ref{sol:Bphi}).
Another condition is that $B_\theta(r=R_c)=0$. This condition is written
as
\begin{equation}
 \tan (\lambda \eta) = \frac{\lambda \eta}{1-(\lambda \eta)^2}.
  \label{eq:lambdarelation}
\end{equation}
Equation (\ref{eq:lambdarelation}) determines $\lambda$ and then the
parameter $H_0$ is obtained from equation (\ref{eq:H0relation}).
Note that the equation (\ref{eq:lambdarelation})
has an infinite number of roots (see Fig.~4 in
\citealt{1984ApJ...281..392L}). 
The first root for $\eta > 0$ arises at $\lambda \eta = \lambda
\eta_1\simeq 2.7$ and the second root does at $\lambda \eta
= \lambda \eta_2\simeq 6.1$. 
The first root corresponds to the position of the centre of the flux
ropes. We take the second root as the contact surface, i.e.,
$\eta_c=\eta_2$ ($z_1$ in Fig.~4 of 
\citealt{1984ApJ...281..392L}) throughout this paper.
The parameter
$H_0$ is then determined from equation (\ref{eq:H0relation}) as
$H_0\simeq 0.81$.

Another constraint on the parameter comes from pressure balance across the
contact discontinuity. The pressure $P_o$
consists of three component of the pressure, $P_A$, $P_Q$ and
$P_i$. $P_A$ and $P_Q$ are exactly zero at $r=R_c$ since
$\tilde{A}|_{\eta=\eta_c}=\partial \tilde{A}/\partial \eta
|_{\eta=\eta_c} =0$ .
$P_i$ should be smoothly connected with $P_s$ at the contact
surface. From this condition, the parameter $P_0$ is expressed as
\begin{equation}
 P_0=\frac{(1-\xi\eta_c^2)^2}{4\nu^3}\left(\frac{GM}{\eta_c}-\frac{\mu}{\sqrt{1-\xi\eta_c^2}}\right)^4.\label{sol:P0}
\end{equation}
Here we used equations (\ref{sol:Pi}) and (\ref{sol:Ps}).

The remaining parameters are $\xi$, $\eta_c$, $R_0$, $\alpha_n$, $n$, $\nu$,
and $\mu$, where $\xi$ denotes the scaling of time and radius and $\eta_c$
describes the velocity of the contact surface. 
Equation (\ref{sol:Rs}) determines $R_0$ by prescribing the shock radius when
self-similar expansion starts.
The twist injection at the central star determines the 
amplitude $\alpha_n$ and the Fourier mode number $n$ of the
toroidal magnetic fields. 
A constant $\nu$ which appears in equation (\ref{eq:adiabatic}) denotes
the entropy of the isotropic plasma in the outflow region
$r \lid R_c$. 
The density at the contact surface when self-similar expansion starts
determines the constant $\mu$ which appears in equation (\ref{sol:Di}).

\begin{figure}
 \begin{tabular}{ccc}
  \begin{minipage}{0.49\hsize}
   \center
   \includegraphics[width=4.8cm]{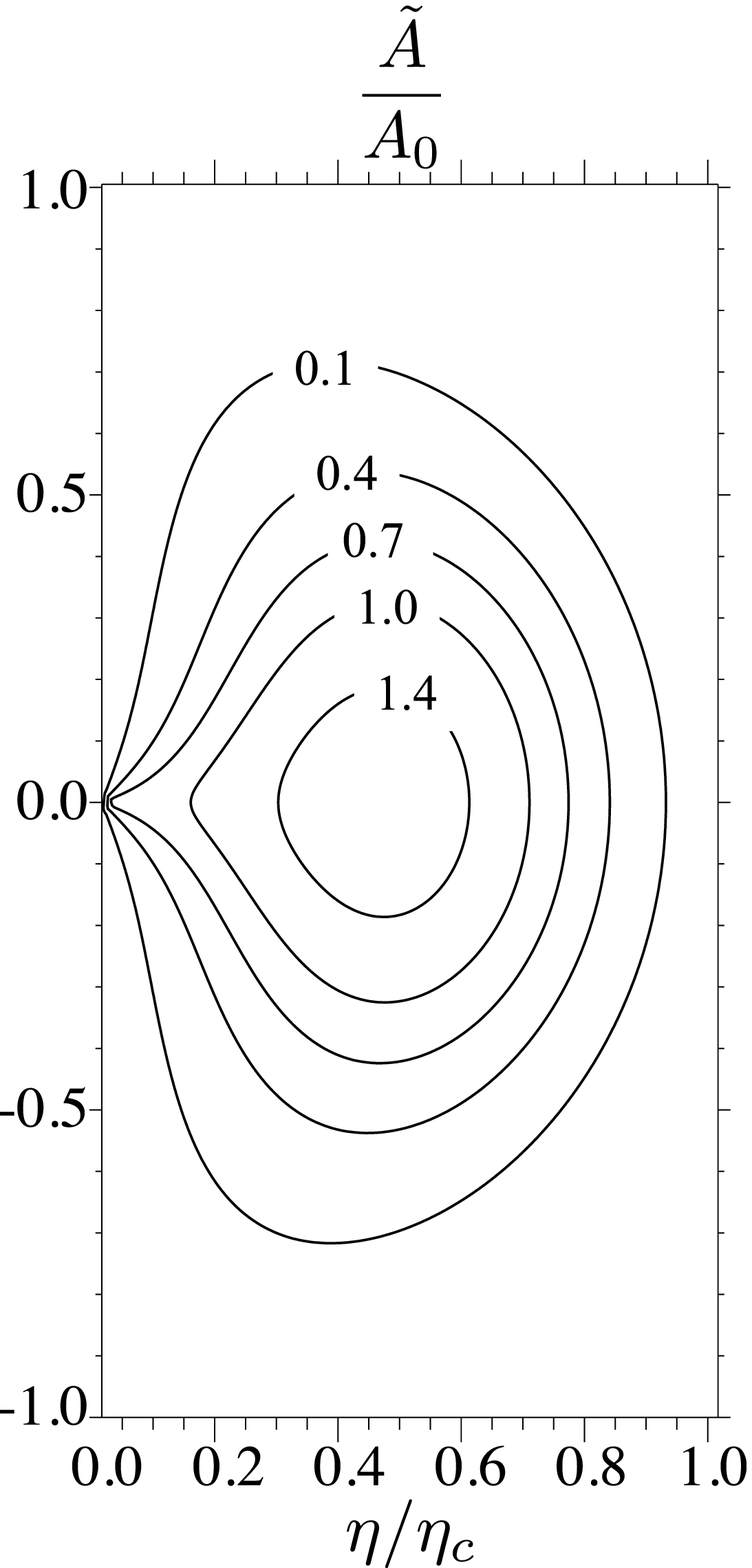}
  \end{minipage}
  \hspace*{1mm}
  \begin{minipage}{0.49\hsize}
   \center
   \includegraphics[width=4.8cm]{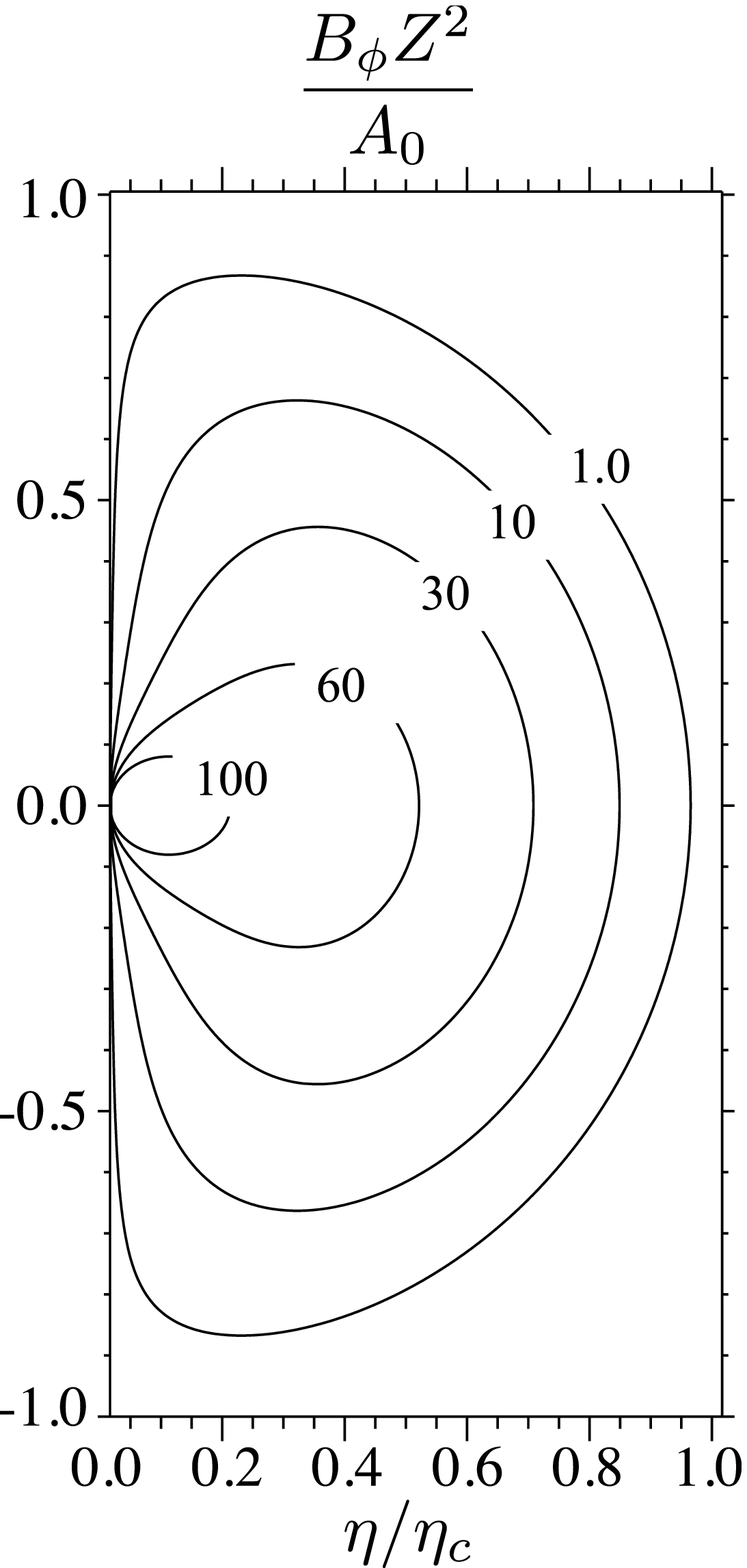}
  \end{minipage}
 \end{tabular}
 \caption{Contour plots of the magnetic flux $\tilde{A}$ (left) and the
 toroidal magnetic fields. The
 contact surface is situated at $\eta/\eta_c=1$. The parameters are taken as
 $\eta_c=\lambda$, $\xi=1$ and $n=1$. }
 \label{fig:mag}
\end{figure}

\section{Physical Properties of the Self-similar
 Explosions}\label{Property}
First, we concentrate on the structure of the magnetic loops in $r\lid
R_c$. Fig.~\ref{fig:mag} shows contours of the magnetic
flux $\tilde{A}$ (left) and the toroidal magnetic fields (right) in
the $\eta-\theta$ plane.  The parameters are $\eta_c=\lambda$ and $\xi=1$.
The poloidal mode number of the toroidal magnetic fields is $n=1$.
We can see the flux rope structures emerging inside the
expanding magnetic loops. The centre of the flux ropes is situated at
$\eta=\eta_1$
($\eta_1 \simeq 0.44 \eta_c$). 
Since the magnetic fields have both poloidal and toroidal components,
they describe the twisted flux ropes. The flux ropes rise in
the $+r$ direction with time. In the limit of $t \gg r$, the magnetic
fields are represented as 
\begin{equation}
 \lim_{t\rightarrow \infty}\bmath B = \frac{2 A_0
  H_0}{r^2}\cos\theta~\bmath e_r, \label{sol:Blimit}
\end{equation}
from equations (\ref{sol:Br}) - (\ref{sol:Bphi}). The toroidal magnetic
fluxes are diluted by the expansion according to the flux conservation
equation. 
The configuration of the
magnetic fields approaches that of the split monopole.

\begin{figure}
 \begin{tabular}{ccc}
  \begin{minipage}{0.49\hsize}
   \center
   \includegraphics[width=4.8cm]{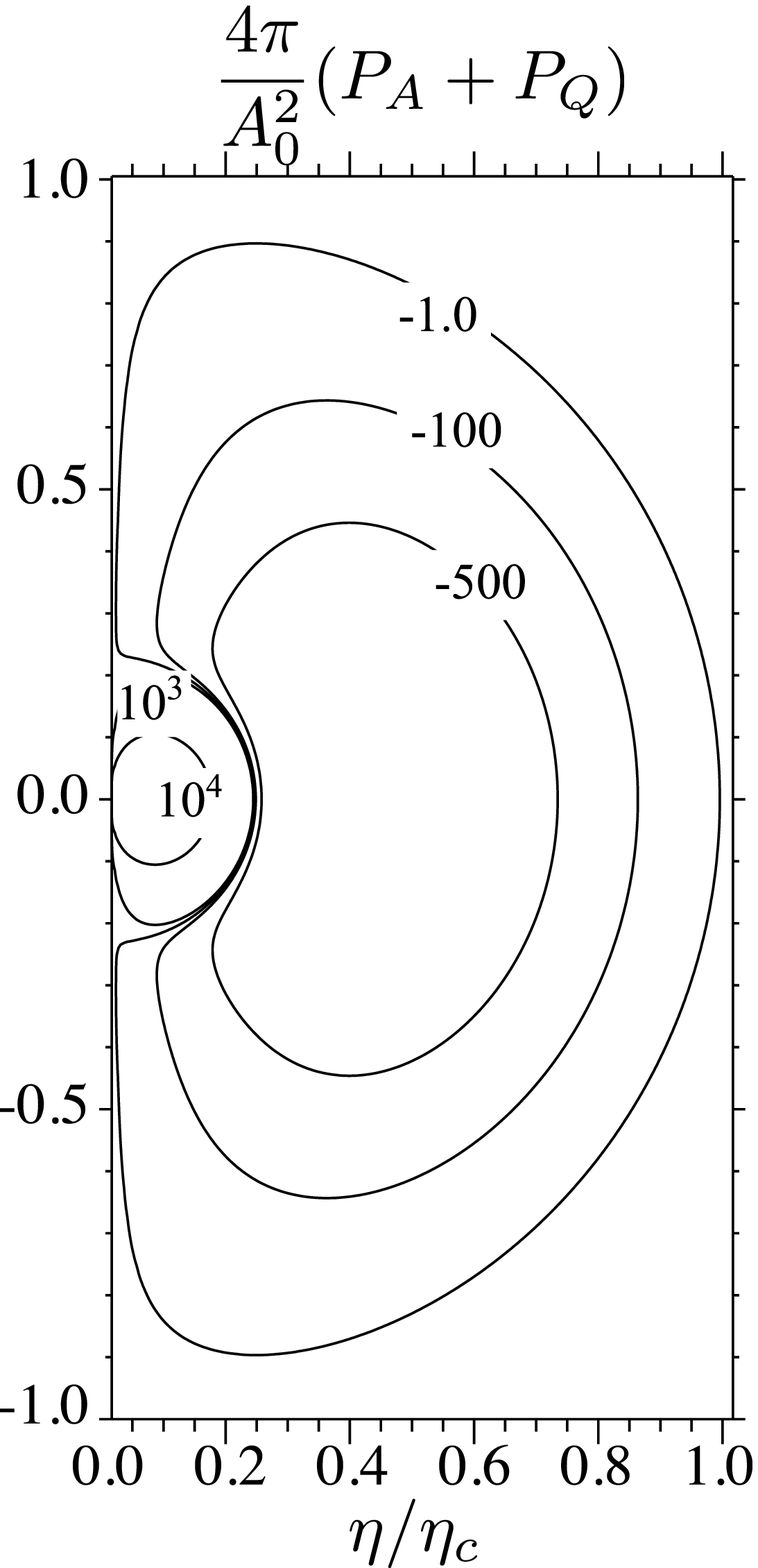}
  \end{minipage}
  \hspace*{1mm}
  \begin{minipage}{0.49\hsize}
   \center
   \includegraphics[width=4.8cm]{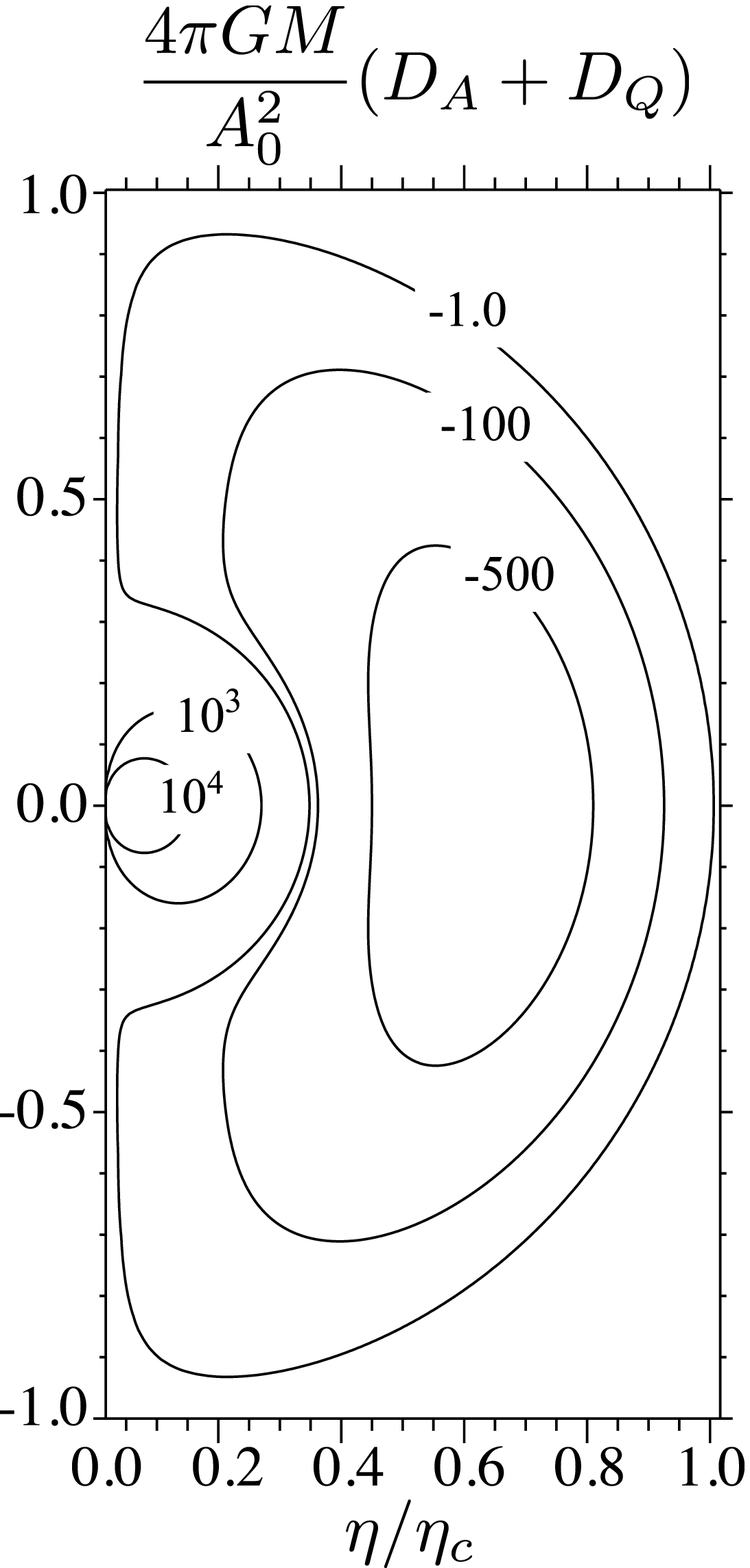}
  \end{minipage}
 \end{tabular}
 \caption{Contour plots of the pressure (left) and the density (right).
 The contribution from the isotropic plasma is subtracted.
The contact surface is situated at $\eta/\eta_c=1$. The parameters are taken as
 $\eta_c=\lambda$, $\xi=1$, $n=1$, and $\mu=0$. }
 \label{fig:pol}
\end{figure}

Fig.~\ref{fig:pol} shows contours of the gas pressure (left) and the
density (right) in the $\eta - \theta$ plane. Contributions from the
isotropic plasma, $P_i$ and $D_i$, are subtracted, so that the pressure
and the density can be negative in these panels. The set of the
parameters is the same as those in Fig.~\ref{fig:mag}. 
Accompanying the flux ropes, low density voids are generated.
 The toroidal magnetic fields can create such voids. 
As mentioned at the end of \S~\ref{derivation}, since the
force balance is attained in our self-similar solutions, 
the Lorentz
force by the poloidal magnetic
fields should balance with the gas pressure gradient force and the
toroidal magnetic pressure gradient force. This
indicates that the gas pressure decreases as the toroidal
magnetic pressure increases.
Such voids exert the buoyancy force on the
plasma in radial direction. 
Fig.~\ref{fig:pol} shows that the pressure
gradient force ahead of the voids balances the buoyancy force.

Next we consider the isotropic part of the outflowing plasma expressed
in equations (\ref{sol:Pi}) and (\ref{sol:Di}). When the
constant $\mu$ is exactly zero, the plasma distribution reduces to that in
hydrostatic states. The scaling comes from the
assumption of adiabatic expansions with the polytropic index of $\Gamma=4/3$.
When $\mu$ is not zero, the plasma distribution differs from that in the
hydrostatic states. When the flow speed is non-relativistic, i.e.,
$\sqrt{\xi} \eta \ll  1$, the solutions reduce to those in non-relativistic MHD obtained by
\cite{1984ApJ...281..392L}. Since the enthalpy contributes to the plasma
inertia, the correction term $(1-\xi \eta^2)^{-1/2}$ arises in the relativistic
MHD. 
The plasma tends to be hydrostatic since $\sqrt{\xi}\eta \ll 1$
when $t\gg r$.

\begin{figure}
 \begin{tabular}{ccc}
  \begin{minipage}{0.49\hsize}
  \includegraphics[width=8cm]{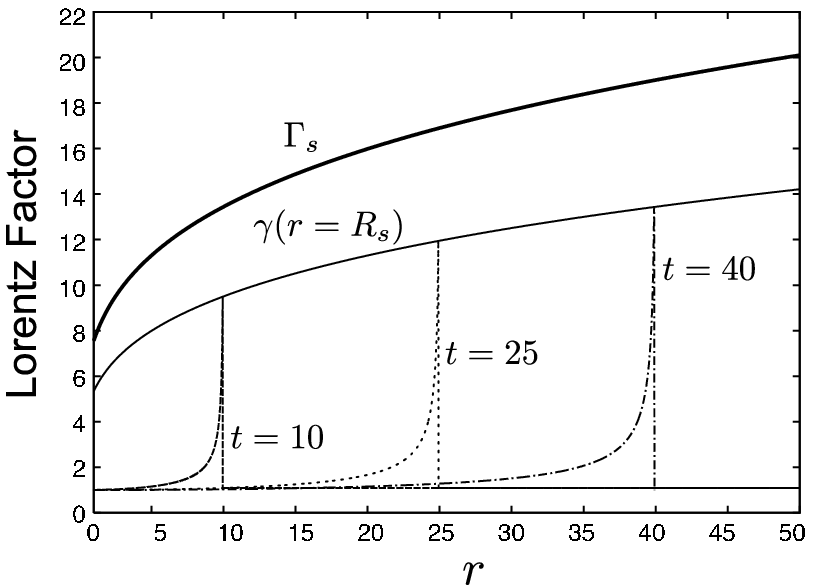}
  \caption{Radial profile of the outflow Lorentz factor. Dashed, dotted,
  and dot-dashed curves show the outflow Lorentz factor at $t=10, 25, 40$,
  respectively. Thick solid curve shows the shock Lorentz factor, while
  thin solid curve does the maximum Lorentz factor of the outflow. 
   The
   value of the parameter $R_0$ is taken to be $R_0=1.78\times 10^{-4}$,
   which corresponds to $\gamma(t=5)=8$.}
  \label{fig:shockvel}
  \end{minipage}
  \hspace*{1mm}
  \begin{minipage}{0.49\hsize}
  \includegraphics[width=8cm]{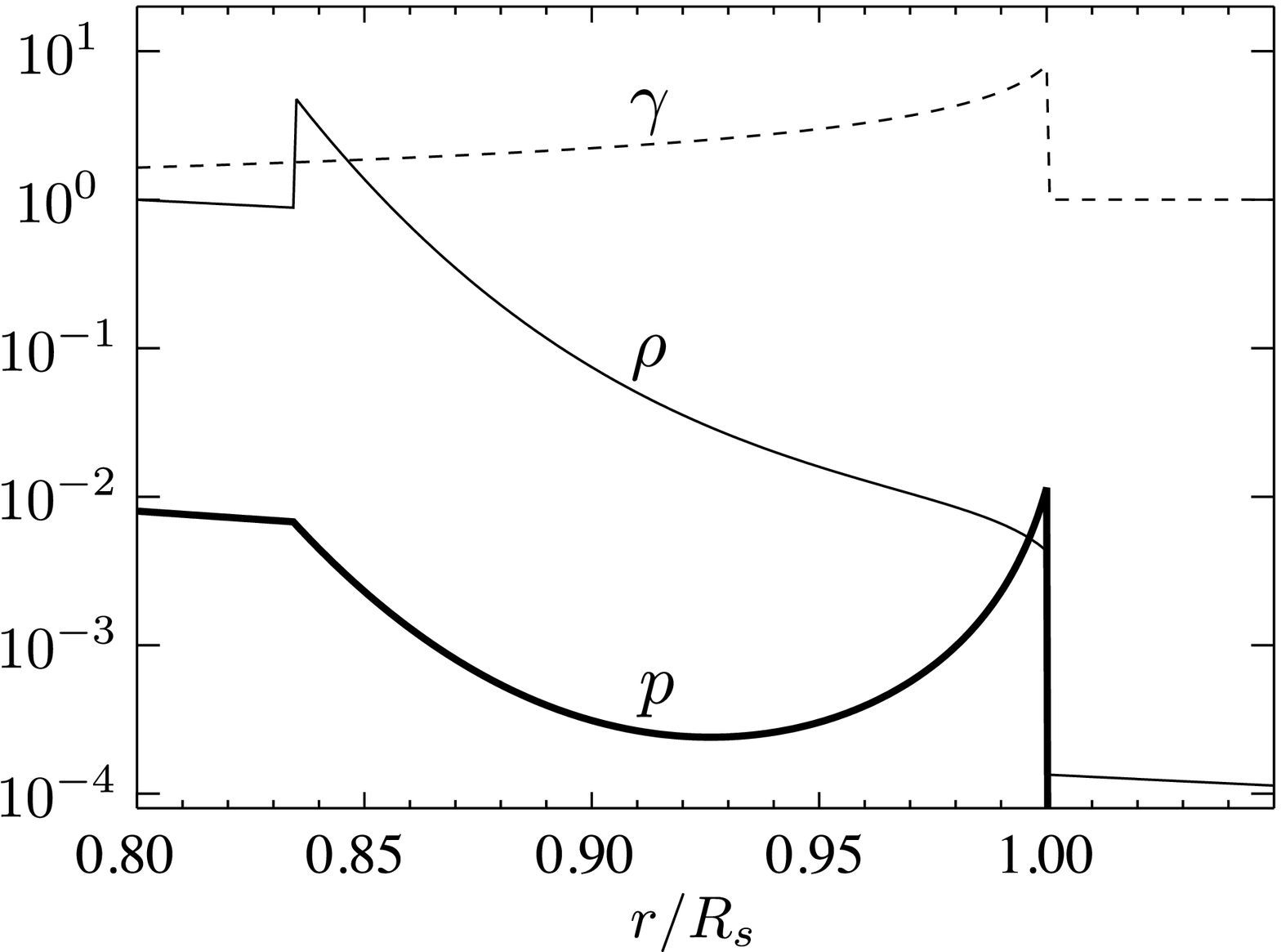}
   \caption{Radial profiles of the plasma pressure (thick solid curve),
   the plasma density (thin solid curve) and the Lorentz factor (dashed
   curve). Radius is normalized by the shock
   radius $R_s$. The contact discontinuity is situated at $R_c/R_s\simeq
   0.834$. We take $\Gamma_s=8\sqrt{2}$ and $\mu=0$.}
  \label{fig:hydro-r}
  \end{minipage}
 \end{tabular}
\end{figure}

In front of the outflow region, the ambient plasma is compressed by the
shocks at $r=R_s$ and is accumulated in the post shock region ($R_c\lid r
\lid R_s$). The contact
surface at $R_c=\sqrt{\xi}\eta_c t$ divides the outflowing plasma from
the shocked plasma. Since we assumed
that the postshock gas evolves self-similarly according to the same basic
equations for the outflow region, the contact surface has the constant
velocity $v_c=\sqrt{\xi}\eta_c$. 
The expansion speed of the shock radius $\mathrm{d}R_s/\mathrm{d}t$
is, however, not constant but it increases with time.
The time evolution of the shock radius
is expressed by equation (\ref{sol:Rs}).
Fig.~\ref{fig:shockvel} shows the radial profile of the
outflow Lorentz factor. Dashed, dotted, and dot-dashed curves denote the Lorentz
factor at $t=10, 25, 40$, respectively. Thick curve denotes the time
evolution of the shock
Lorentz factor $\Gamma_s$, while thin solid curve does the time profile
of the outflow Lorentz factor at the shocks ($\gamma|_{r=R_s}$).  
The value of the parameter $R_0$ is $R_0=1.78\times 10^{-4}$, which
corresponds to $\gamma(t=5)=8$.
The shock Lorentz factor is larger than the outflow Lorentz factor at
$r=R_s$ by factor $\sqrt{2}$, as expected from relativistic strong gas
dynamical shocks (see equation \ref{eq:RHg}). The undisturbed plasma is
abruptly heated up by the shocks. 
Plasma velocity suddenly becomes zero ahead of the shocks $(r>R_s)$
where the undisturbed plasma exists.

As the shock propagates in the undisturbed plasma, the shock surface is
accelerated. 
The shock Lorentz factor and the shock radius can be expressed as
\begin{equation}
 \Gamma_s \simeq \left[\frac{(1+\sqrt{2})^{\sqrt{2}}}{6}\right]^{1/4}
  \left(\frac{t}{R_0}\right)^{1/4}\simeq 0.87
  \left(\frac{t}{R_0}\right)^{1/4}, \label{sol:appLorentz}
\end{equation}
\begin{equation}
 R_s\simeq t,\label{sol:appRs}
\end{equation}
respectively.
Here we approximate $v\simeq 1$ to obtain the first equality in equation
(\ref{sol:appLorentz}). 
While the shock surface moves with almost constant speed $\simeq 1$, 
the shock Lorentz factor increases with time with the power law index of
$0.25$. 
This result comes from the density profile of the undisturbed plasma. As
shown later, the undisturbed plasma density decreases with radius as
$r^{-3.5}$ (see equation
\ref{sol:rho0-apx}). \cite{1979ApJ...233..831S} showed that
when $\rho\propto r^{-\delta}$, the flow is accelerated when $\delta>3$.
Although we take into account the gravity
from the central star, which is not included in
\cite{1979ApJ...233..831S}, the gravitational force is smaller than the
pressure gradient force in this region. Thus their relation can be adopted in
our analysis. The outflow is gradually accelerated when $\delta\simeq 3.5$.

Next we consider the radial profile of the plasma pressure and the
density. 
Fig.~\ref{fig:hydro-r} shows the radial profile of the pressure $p$
(thick solid curve), the density $\rho$ (thin solid curve) and the
outflow Lorentz factor (dashed curve) on the equatorial plane. 
The horizontal axis denotes the radius normalized by the shock radius $R_s$. 
The shock Lorentz factor is taken as $\Gamma_s=8\sqrt{2}$. We take
$\mu=0$ for simplicity. The other parameters are $\xi=1$, $R_0=1.3\times
10^{-4}$, $\alpha_n=0$, and $A_0=0$. The contact surface is situated at
$R_{c}\simeq 0.834R_s$. 
Behind the contact surface, the gas pressure and the density decrease
with radius as $\propto r^{-4}$ and $\propto r^{-3}$, respectively.
The density jump appears at $r=R_{c}$, while the pressure is
continuous (contact surface). The swept-up ambient plasma is accumulated
in region $R_c\lid r \lid R_s$. The plasma density in the postshock
region ($0.834<r/R_s<1$) is
larger than that in the outflow region. The
strong pressure gradient force in this postshock region pushes the
plasma in $+r$ direction. 
The pressure gradient force balances with the inertia of the accumulated
plasma. Thus, the plasma flows toward $+r$ direction with the
inertial velocity.

Ahead the contact discontinuity, a strong shock appears at
$r=R_s$.
We assumed the strong shock and neglected the ambient plasma pressure.
The plasma is abruptly heated up by the shock.
The ambient plasma density jump also appears at the shocks. The density of the shocked
gas is larger than that of the
ambient plasma by factor $32$ for $\Gamma_s=8\sqrt{2}$ as expected from
the relativistic Rankin-Hugoniot relations (see equations \ref{eq:RHr}
and \ref{eq:RHg}). 
The density of the undisturbed plasma is described by equation
(\ref{sol:rho0}). From this equation, the density can be approximately
represented by the power law of $r$ as
\begin{equation}
 \rho(r)\simeq \frac{3 P_0}{8\sqrt{R_0}\xi^2}\sqrt{\frac{(1+\sqrt{2})^{\sqrt{2}}}{6}}K(1)r^{-\frac{7}{2}}.\label{sol:rho0-apx}
\end{equation}
Here we approximate $v_r(r=R_s)\simeq 1$ for relativistic flows. The
density decreases with radius with the power law index of $-3.5$.
The ambient plasma density decreases slightly faster than that inside
the shocks ($\propto r^{-3}$). The shock Lorentz factor thus increases
with radius \citep{1979ApJ...233..831S}.

Next we consider the total energy contained within the spherical
surface $R_s$. Let $\mathcal{E}$ be the total energy in $r<R_s$. 
As shown in \cite{2009MNRAS.394..547T}, the virial theorem can be
applied for the relativistic inertial flow: 
\begin{equation}
 \mathcal{E}=\mathcal{K}-(3\Gamma-4)U_\mathrm{th} 
+\int \frac{\partial \bmath{S}\cdot \bmath r}{\partial t}d V 
+\int p \bmath r \cdot d
  \bmath{\mathcal{A}}
  -\frac{1}{8\pi}\int 
  \left\{2\left[(\bmath r \cdot \bmath E)(\bmath E\cdot
	 d \bmath{\mathcal{A}})+(\bmath r \cdot \bmath B)(\bmath B\cdot
	 d \bmath{\mathcal{A}})
\right]-(\bmath E^2+\bmath B^2)(\bmath r\cdot d \bmath{\mathcal{A}})\right\}, \label{eq:virial}
\end{equation}
where 
\begin{equation}
 \mathcal{K}=\int dV \rho\gamma^2,\label{eq:kinetic}
\end{equation}
\begin{equation}
 U_\mathrm{th}=\int dV \frac{p}{\Gamma-1}, \label{eq:thermal}
\end{equation}
and $\bmath{S}$ shows the Poynting flux. Here $\bmath{\mathcal{A}}$ denotes
expanding spherical surface at $r=R_s(t)$. The third, fourth, and fifth terms
of the right hand side of equation (\ref{eq:virial}) represent the
Poynting flux, the work done by the gas
pressure, the work done by the Maxwell stress, respectively.

Let us evaluate the non-kinetic part of the energy,
$\mathcal{E'}=\mathcal{E}-\mathcal{K}$. The second term in the right
hand side of equation (\ref{eq:virial}) is zero because the
polytropic gas index is $\Gamma=4/3$. 
The fifth term of (\ref{eq:virial}) is zero since the
electromagnetic fields vanish at $r=R_s$.
The third term also becomes zero after the straightforward calculations. 
Thus the non-kinetic part of the energy can be evaluated as
\begin{equation}
 \mathcal{E}'(t)=\left.\frac{8\pi}{3}\rho_0\Gamma_s^2R_s^3\right|_{r=R_s}=\left.
		  \frac{4\pi}{3}\rho \gamma R_s^3\right| _{r=R_s}.\label{sol:nonE}
\end{equation}
Here we used equations (\ref{eq:RHp}) and (\ref{eq:RHg}).
The energy $\mathcal{E}'$ does not depend on the amplitudes of the
magnetic fields. As shown in \S~\ref{expsol}, the plasma density
(pressure) consists of three parts, $D_A$, $D_Q$ and $D_i$ ($P_A$, $P_Q$
and $P_i$). 
While the components $D_A$ and $D_Q$ depend on the magnetic field
strength, $D_i$ is independent of them. $D_A$ and $D_Q$ do not
contribute to the non-kinetic part of the energy from equation
(\ref{eq:virial})
since they are exactly zero at $r=R_c$. 
This indicates that the plasma interacting with the magnetic
fields is in marginally stable state.
\cite{1982ApJ...254..796L} showed that the inertial
flow with the polytropic index $\Gamma=4/3$ represents the marginally
stable state in non-relativistic MHD. This situation is also valid for
the relativistic MHD. Total energy contained inside the shocks is thus
independent of the strength of the magnetic fields.

By substituting equations (\ref{sol:appLorentz}) and
(\ref{sol:rho0-apx}) into equation (\ref{sol:nonE}), the time evolution of
the energy $\mathcal{E}'$ is written as
\begin{equation}
 \mathcal{E}'=\frac{(1+\sqrt{2})\pi}{6}\frac{K(1)}{\xi}\frac{P_0}{R_0},
  \label{sol:Edash}
\end{equation}
where we used equations (\ref{sol:appLorentz}), (\ref{sol:appRs}), and
(\ref{sol:rho0-apx}).  
Note that the non-kinetic part of the energy
$\mathcal{E}'$ is positive.
This means that plasma speed exceeds the
escape velocity determined by the gravitational potential.

The kinetic energy $\mathcal{K}$ given in equation (\ref{eq:kinetic}) is
expressed as
\begin{equation}
 \mathcal{K}=2\pi \int d\theta \sin\theta \int d \eta~ \frac{D(\eta, \theta)}{\sqrt{1-\eta^2}},
  \label{eq:kinetic2}
\end{equation}
where we used equations (\ref{eq:rho}) and (\ref{sol:v}). Note that
equation (\ref{eq:kinetic2}) depend on time
through $\eta$.
When we integrate inside the sphere with radius $r=R_s$, the
integration is carried out in $[0,\eta_s]$ in the self-similar space. 
Since the flow is relativistic, i.e., $v_r\simeq 1$, $\eta_s\simeq
1/\sqrt{\xi}$ is constant with time. Thus both non-kinetic and kinetic
energies are constant with time.
Strictly speaking, the total energy should increase with time  
since the shock surface sweeps up the ambient plasma. The rest mass energy of
the swept-up plasma contributes to the increase in the total energy. 
This energy is, however, negligible because we assume the strong
shocks. From equation (\ref{eq:RHp}), the rest mass energy density of
the undisturbed ambient plasma is
smaller than the kinetic energy density of the shocked gas by factor
$\Gamma_s^4$ and negligibly small (Note that the relations \ref{eq:RHp}
- \ref{eq:RHg} are correct with the accuracy of $\mathcal{O}(1/\Gamma_\mathrm{s}^2)$). 
For the same reason, $\mathcal{E}'$ is also independent of time with the
accuracy of $\mathcal{O}(1/\Gamma^2)$. 
Using these facts, 
the shock Lorentz factor is expressed as
\begin{equation}
 \Gamma_s =
  \sqrt{\frac{3}{8\pi}}\mathcal{E'}^\frac{1}{2}\rho_0^{-\frac{1}{2}}(r=R_s)
  R_s^{-\frac{3}{2}}, \label{eq:Gs}
\end{equation}
from equation (\ref{sol:nonE}). This result is equivalent with
equation (16) in \cite{1979ApJ...233..831S}. Although our solutions
include the magnetic fields and the gravity from the central star, they
do not contribute to the total energy because the system is in a
marginally stable state. Thus only the hydrodynamical (isotropic) part
contributes to the the total energy. 
By inserting equations
(\ref{sol:appRs}) and (\ref{sol:rho0-apx}) into this equation, we obtain
$\Gamma_s\propto r^{1/4}$ again (see equation \ref{sol:appLorentz}). The shock is thus accelerated when it
propagates in the ambient plasma.

\section{Numerical simulations}\label{numerical}
In this section, we show results of relativistic MHD simulations to
study the stability of our solutions. For this purpose, we use the
analytical solutions as the initial conditions of the numerical calculations.  
The relativistic MHD equations are solved in two dimensions using polar
coordinates ($r$, $\theta$). We assume axisymmetry ($\partial /(\partial
\phi)=0$). The number of grid points is ($N_r$, $N_\theta$)=(3600, 360) on
the domain of $R_\mathrm{in}\equiv 1 \lid r \lid 50$ in normalized unit
and $0 \lid \theta
\lid \pi$. The grid sizes are $\Delta r=1.39\times 10^{-2}$ and $\Delta
\theta=8.72\times 10^{-3}$.
We use the HLL method \citep{1983siamRev...25..35..61}
to calculate numerical fluxes. We utilize the modified CTU method
\citep{2006MNRAS.368.1040M} to achieve the second order accuracy in
space. In our analytical solutions, the strong shock is expected.
Such a strong shock can induce the numerical oscillations.  
To avoid the problem, we utilize the harmonic mean for smoother prescription
\citep{1977JCoPh..23..263V}.
We use the Constraint-Transport method to satisfy the no monopole
condition. 
We impose the outflow boundary conditions in outer radial boundary at
$r=50$ and the symmetric condition at the axes at $\theta=0, \pi$. 
The inner boundary conditions are imposed at $r=R_\mathrm{in}$ by 
applying the time-dependent analytical solutions.
We solve the whole region covering $\theta \in [0, \pi]$, although our analytical
solution are symmetric at $\theta=\pi/2$. So we can check whether our code 
maintains this symmetry.
The parameters are the initial time $t_0=11.0$, $R_s(t=t_0)=10.8$,
$R_c(t=t_0)=8.83$,
$r_g\equiv  GM = 0.148$,
$\xi=1$, $R_0=2.58\times 10^{-3}$, $\alpha_1=0.1$, $n=1$, $\mu=0$,
and $\rho_i R_i^4/A_0^2=100$. The initial maximum Lorentz
factor ($\gamma_s(t=t_0, r=R_s(t=t_0))$) is $5$. Radius and time are
normalized by the inner radius $R_\mathrm{in}$ and its light crossing time,
respectively.

To obtain the analytical solutions, the ambient plasma pressure
$p_0$ is
not specified because we adopted Ranking-Hugoniot relations for a strong 
shock, so that the ambient pressure is negligible.
However, in order for the ambient plasma to be in hydrostatic
equilibrium, the gas pressure gradient force should balance with the 
gravitational force of the central star. 
Since it is a hard task to reconstruct the self-similar solutions taking 
into account the ambient plasma pressure,
we use the analytical solutions obtained by using approximate
Rankin-Hugoniot relations given in equations 
(\ref{eq:RHp})-(\ref{eq:RHg}) as the initial conditions and the 
ambient pressure is taken initially to be constant, which means
that the ambient plasma is not in hydrostatic equilibrium.
The ambient plasma slowly falls toward the central star due to the
gravity of the central star. The parameters we used in numerical
simulations are taken so that the free fall time $t_{ff}$
is much larger than the dynamical time $t_d$ (typically,
$t_{ff}/t_{d}\simeq 10^{3}$). Thus the free fall motion of the ambient
plasma does not affect the dynamics of the expanding magnetic loops. 
\cite{1992ApJS...80..753S} adopted a different method for this
problem such that the
gravitational force is artificially subtracted in the ambient plasma 
to numerically recover the solutions of the non-relativistic coronal
mass ejection obtained by \cite{1984ApJ...281..392L}. 
We confirmed that the results obtained by the method proposed by them
are consistent qualitatively and quantitatively with those including the
gravity in the ambient plasma. 
The ambient plasma pressure, which is not specified in the
analytical solutions, is taken so small ($p_0=10^{-8}$) that
it does not affect the dynamics of the outflows.

\begin{figure}
 \center
 \includegraphics[width=10cm]{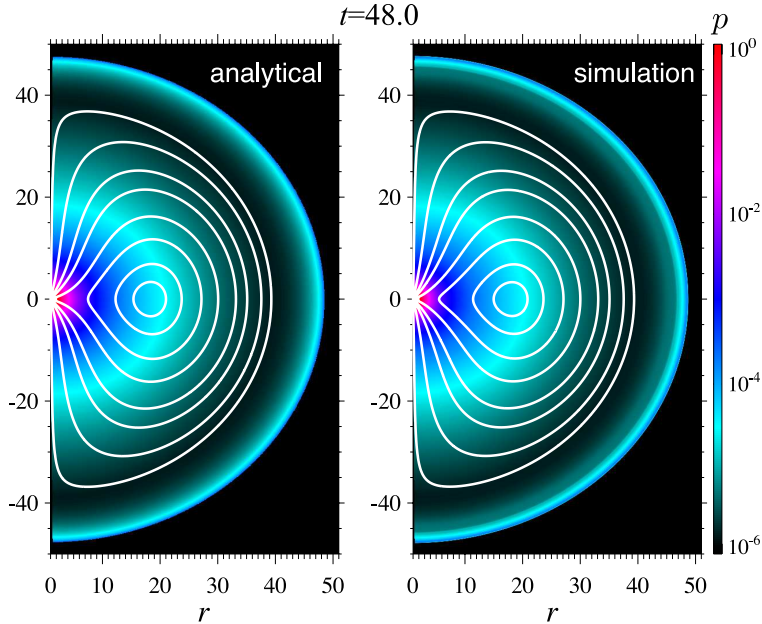} 
 \caption{Colour contour shows the gas pressure profile and curves show
 the magnetic field lines for analytical solutions (left) and
 numerical results (right).}
 \label{fig:pres}
\end{figure}
\begin{figure}
 \center
   \includegraphics[width=10cm]{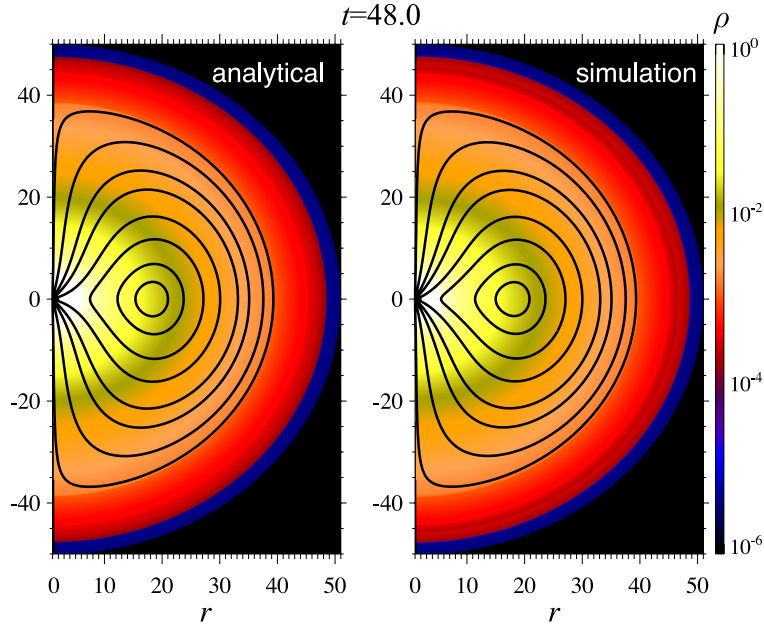}
 \caption{Colour contour shows the density profile and curves show
 the magnetic field lines for analytical solutions (left) and
 numerical results (right).}
 \label{fig:dens}
\end{figure}

Fig.~\ref{fig:pres} shows the pressure profile (colour) and the magnetic
field lines (curves) at $t=48$. The left figure shows analytical solutions
and the right one does numerical results.
Fig.~\ref{fig:dens} shows the density profile (colour). 
The magnetic field lines in Fig.~\ref{fig:pres} and Fig.~\ref{fig:dens} 
are depicted as the isocontours of the flux function. The levels of the 
isocontours are identical in both figures.

As time goes on, the magnetic loops containing the flux ropes expand in
radial direction. The flux ropes carry the toroidal magnetic fields.
The ambient matter inflowing through the shock is compressed and 
accumulated between the contact discontinuity and the shock.
The numerical results excellently recover the analytical solutions.

\begin{figure}
 \begin{tabular}{ccc}
  \begin{minipage}{0.49\hsize}
   \center
   \includegraphics[width=8cm]{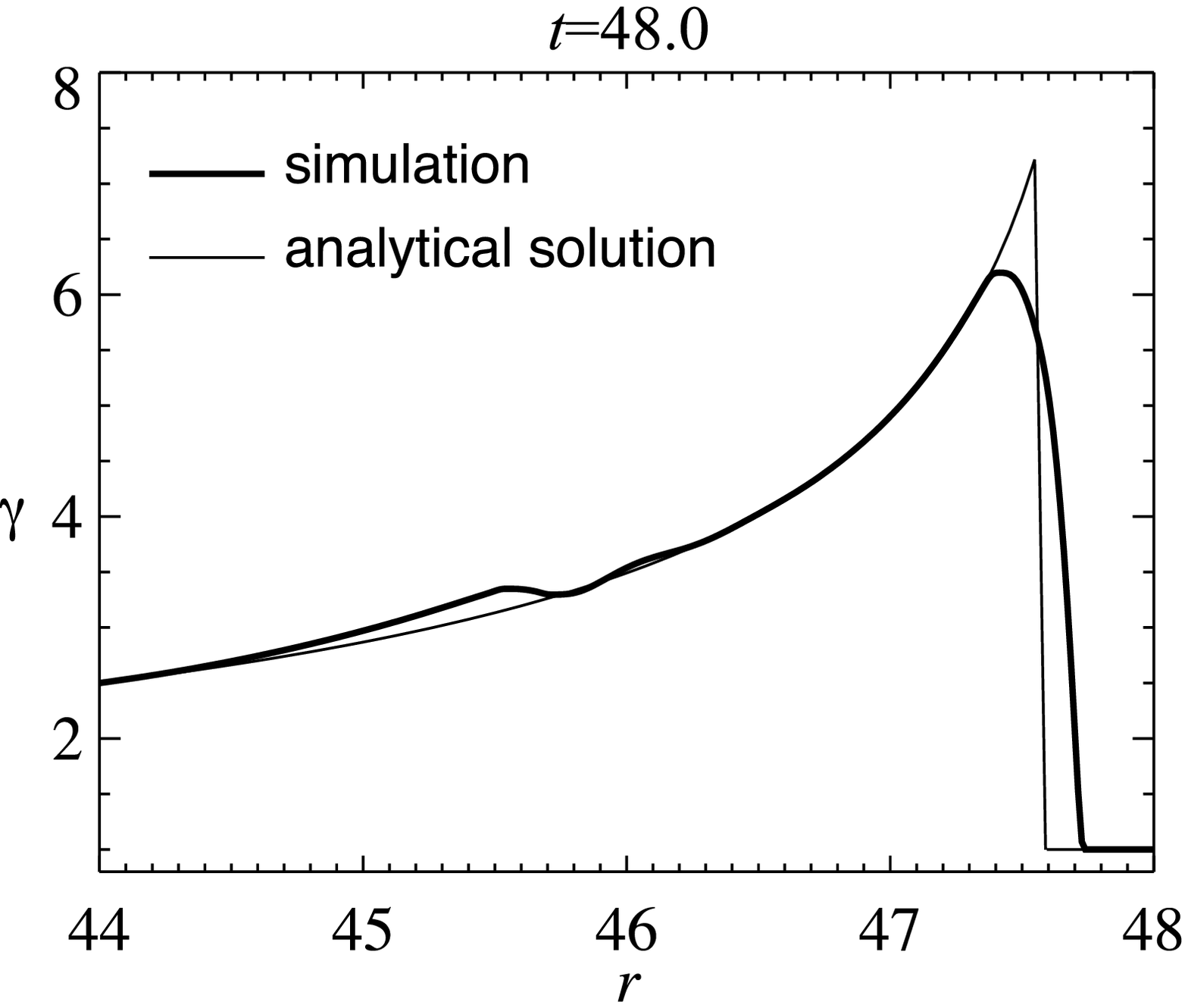}
  \end{minipage}
  \hspace*{1mm}
  \begin{minipage}{0.49\hsize}
   \center
   \includegraphics[width=8cm]{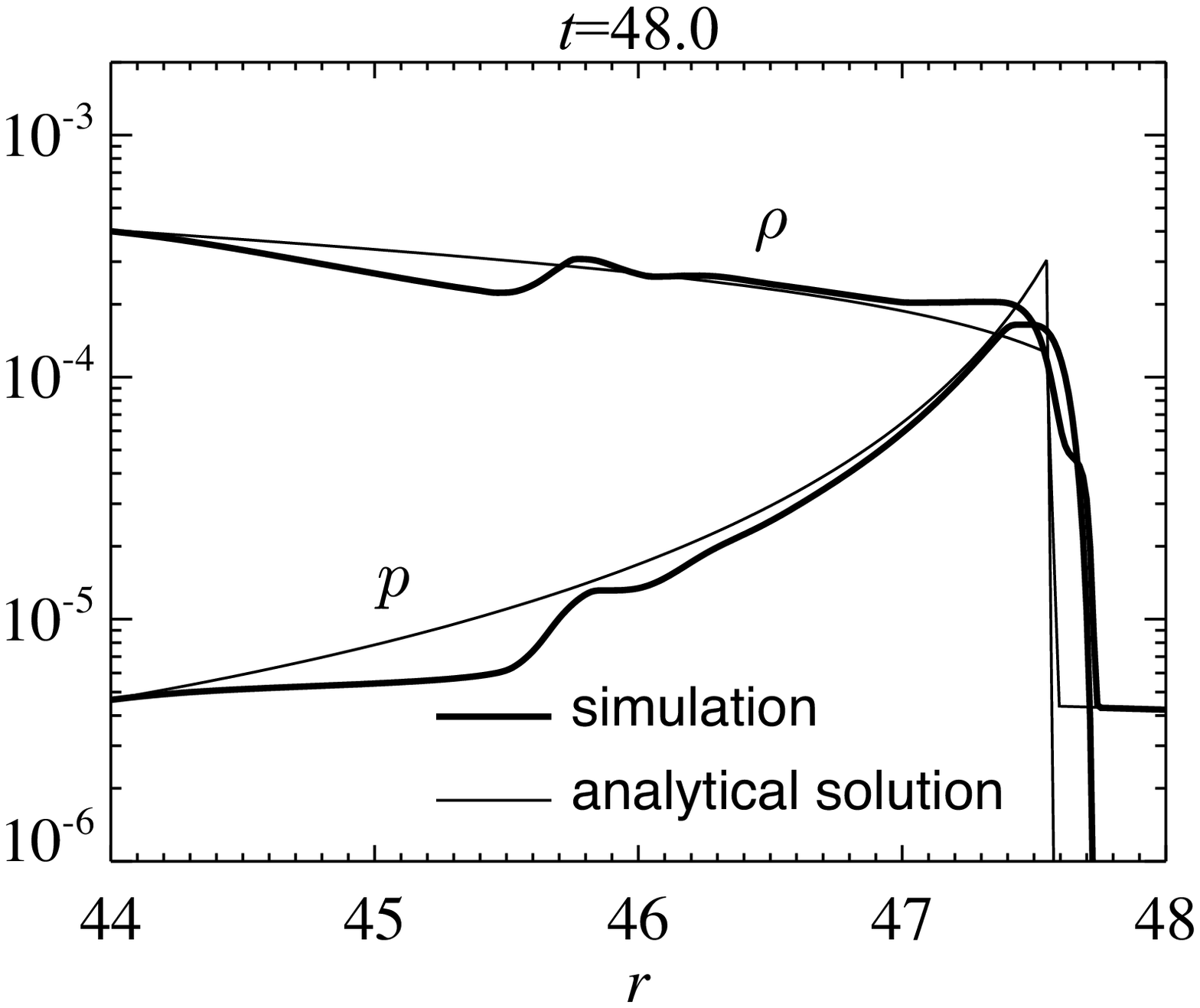}
  \end{minipage}
 \end{tabular}
 \caption{The Lorentz factor (left) and the density and pressure
 profiles (right) on the equatorial plane at $t=48$. Thick curves denote the
 numerical results, while thin curves do analytical solutions.}
 \label{fig:rogm}
\end{figure}

\begin{figure}
 \begin{tabular}{ccc}
  \begin{minipage}{0.49\hsize}  
   \center
   \includegraphics[width=6cm]{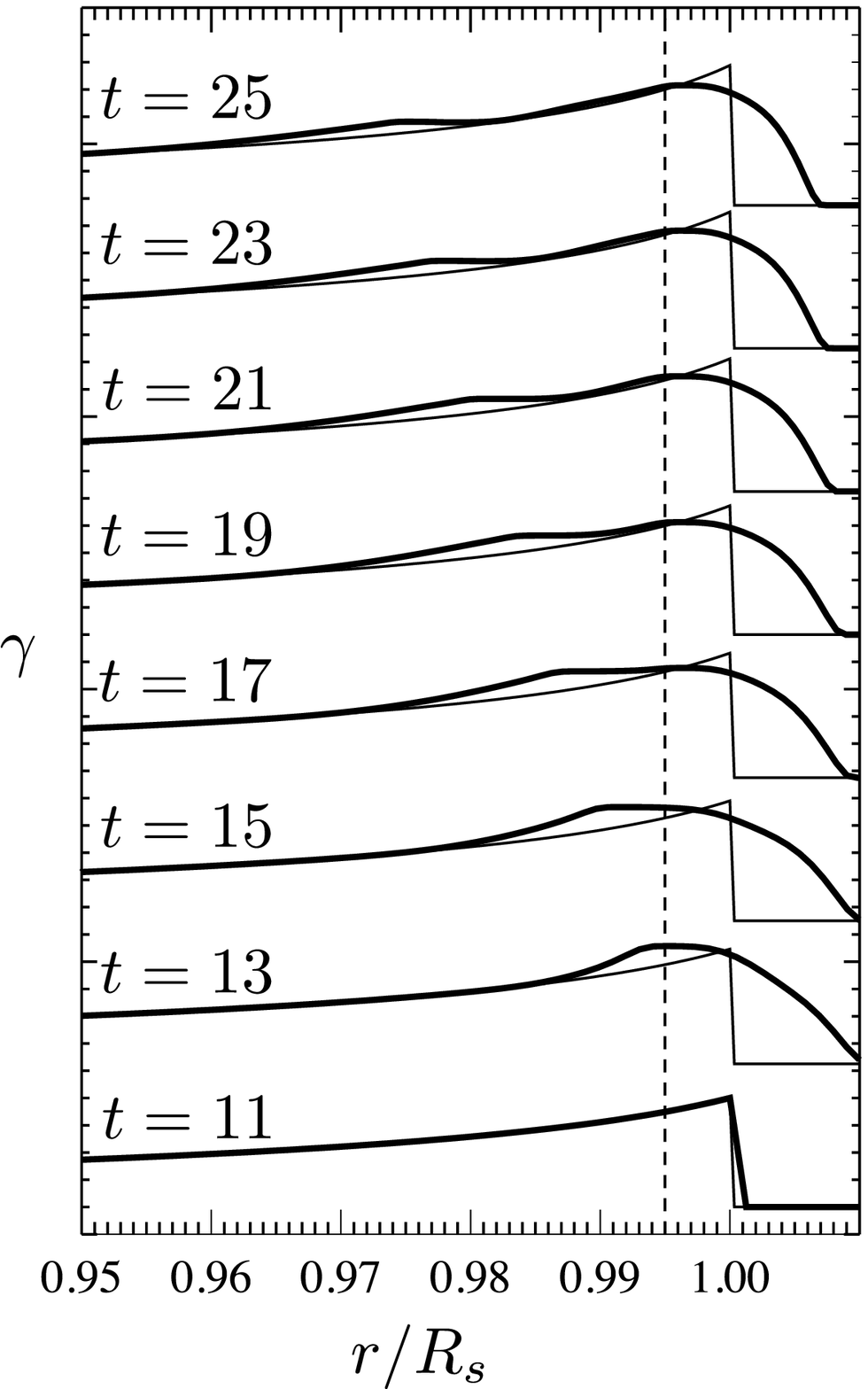}
   \caption{Time evolution of the Lorentz factor on the equatorial plane.
   Thick curves show
   the numerical results, and the thin curves do the analytical
   solutions. Dashed curve shows the reference radius $r=r_*=0.995 R_s$.}
   \label{fig:gamma-evol}
  \end{minipage}
  \hspace*{1mm}
  \begin{minipage}{0.49\hsize}
   \center
   \includegraphics[width=8cm]{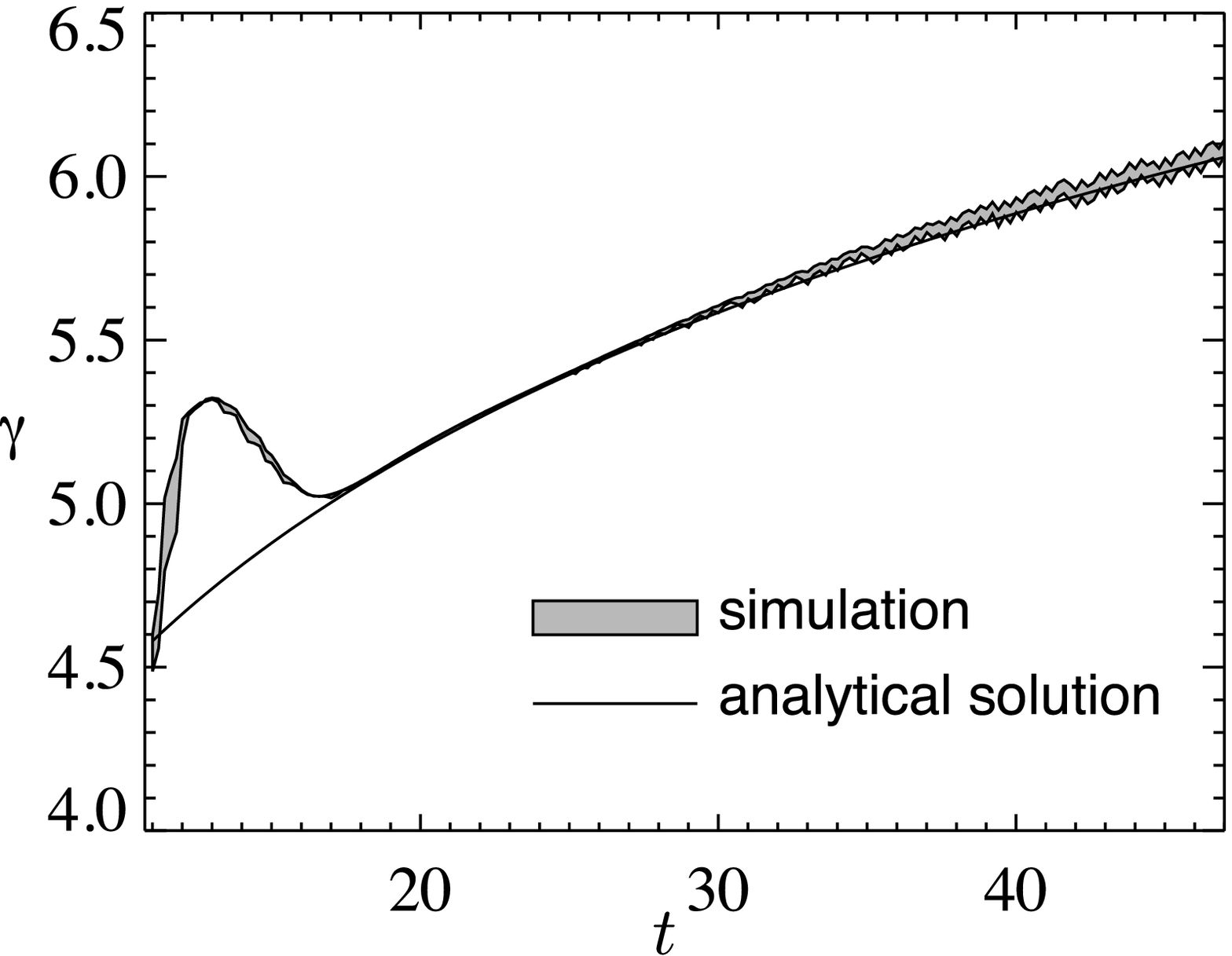}
   \caption{Time evolution of the Lorentz factor at
   $r\equiv r_*=0.995R_s$. Horizontal axis shows the time and
   vertical axis shows the Lorentz factor on the equatorial plane. Solid
   curve shows the analytical solutions.
   Grey contours show the Lorentz factors at the grid points closest to 
   the reference radius $r_*=0.995R_s$.
   }
 \label{fig:gmevol}
  \end{minipage}
 \end{tabular}
\end{figure}
Fig.~\ref{fig:rogm} shows the shock structure on the equatorial plane at
$t=48$. 
Left panel shows the Lorentz factor and the right one does the
density and the pressure. 
Thin solid curves denote the analytical solutions, while
thick solid curves do the numerical results. 
At this time, the maximum Lorentz factor is $\gamma=7.2$ for the
analytical solution. The peak Lorentz factor in the numerical simulation
is, however, $\gamma=6.2$. 
The difference comes from the shock flattening in the
simulations. Note that we
adopt the harmonic mean to evaluate primitive variables on the cell
surface. This method is more diffusive than other interpolation
methods, such as the MUSCL type interpolations. Although we utilize this
method to avoid numerical oscillations at the strong shocks, it 
decreases the peak Lorentz factor. 
Also the density and the pressure
profiles are diffused (right panel of Fig.~\ref{fig:rogm}).
Another reason comes from the assumption of strong shocks adopted to
derive the approximate Rankin-Hugoniot relations (\ref{eq:RHp}) -
(\ref{eq:RHg}). These relations are correct
with an accuracy of $\mathcal{O}(1/\gamma^2)$. Since we take the initial
Lorentz factor at the shock as $\gamma=5$, a few percent error arises
from the approximations. 

Such a numerical diffusion produces the sound waves from the
shocks. 
Fig.~\ref{fig:gamma-evol} shows the radial profile of the Lorentz
factor on the equatorial plane 
from the initial state at $t=t_0=11$ to $t=25$ with
the interval $\delta t = 2$. Thick and thin solid curves show 
numerical results and analytical solutions, respectively, while the dashed
line denotes a reference radius $r_*=0.995R_s$ (see
Fig.~\ref{fig:gmevol}).
After the simulation goes on, the shock is flattened due to the numerical
diffusion, generating the sound waves propagating inward and outward
from the shock.
The compressional waves extract a part of the fluid kinetic
energy from the shock, resulting in the increase in the numerical velocity
behind/ahead of the shock from the analytical solutions.
As the wave propagates away from the shock front, the radial 
profiles approach those of analytical solutions. Note that the 
amplitudes of the inward propagating wave decreases with time because 
the wave conserves the wave energy density $E_w = \rho \gamma^2$. 
Since the density increases inward, the velocity deviation $\delta 
\gamma$ decreases as the wave propagates.

\begin{figure}
   \center
 \includegraphics[width=7.5cm]{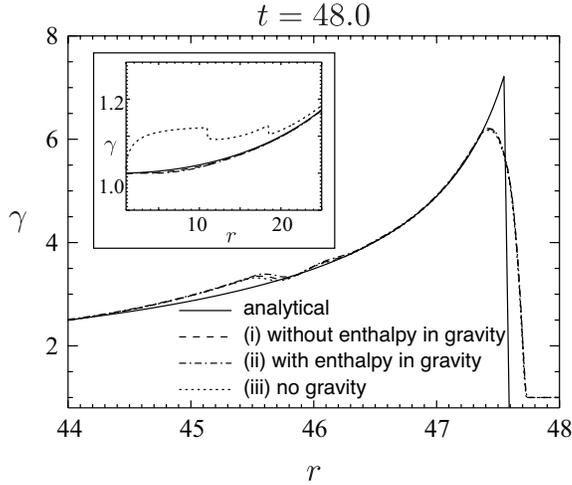}
 \caption{The dependence of the equatorial Lorentz factor at $t=48$ on
 models of the gravity, (i) $-GM\rho \gamma^2/r^2$ (dashed)
 (ii) $-GM\rho h \gamma^2/r^2$ (dash-dotted) and (iii) without
 gravity. Solid curve denotes the analytical self-similar solutions. An
 inset shows solutions in the different range $1 \le r \le 25$.}
 \label{fig:gcomp}
\end{figure}

\begin{figure}
 \begin{tabular}{ccc}
  \begin{minipage}{0.49\hsize}
   \center
   \includegraphics[width=8cm]{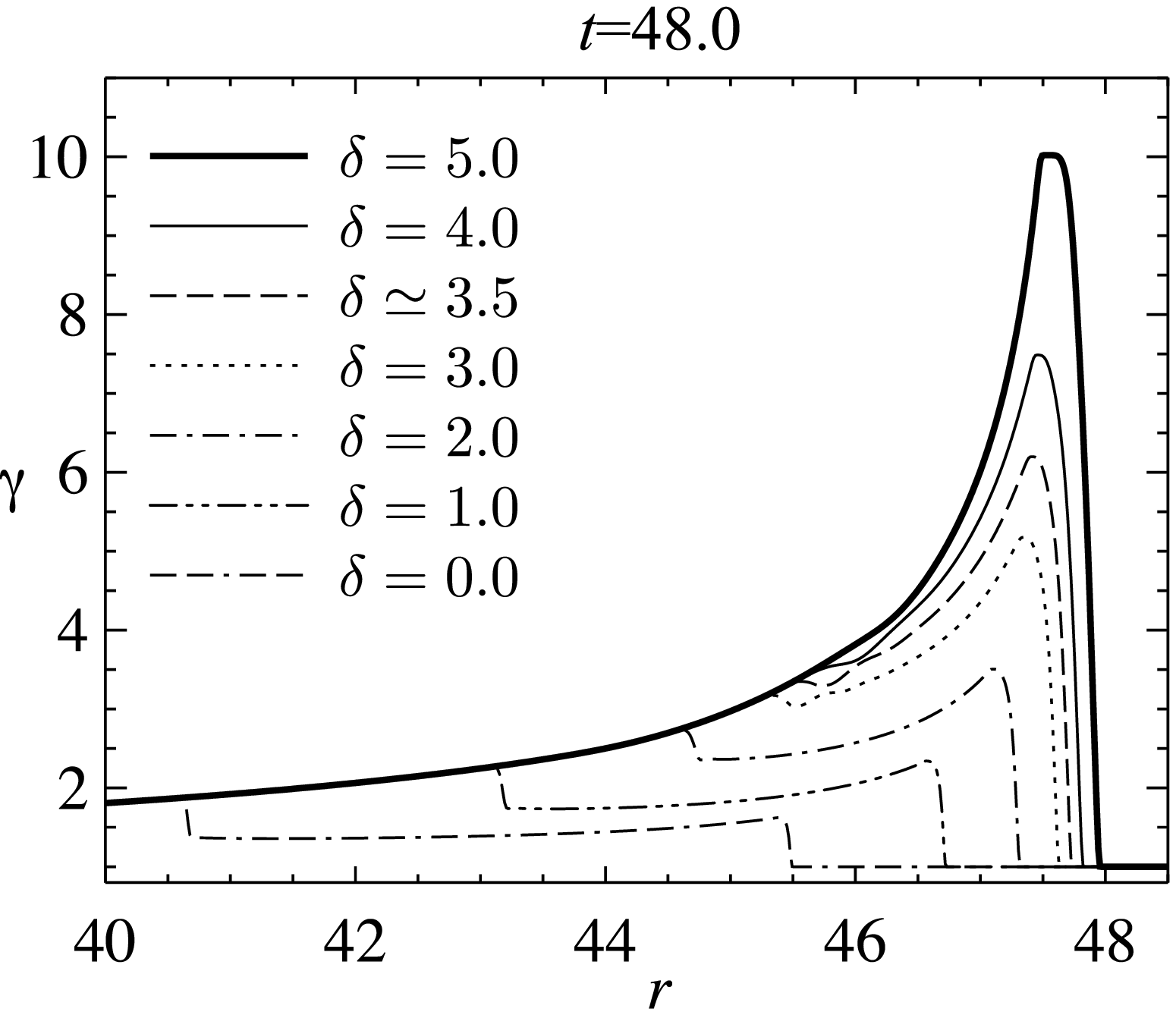}
  \end{minipage}
  \hspace*{1mm}
  \begin{minipage}{0.49\hsize}
   \center
   \includegraphics[width=8cm]{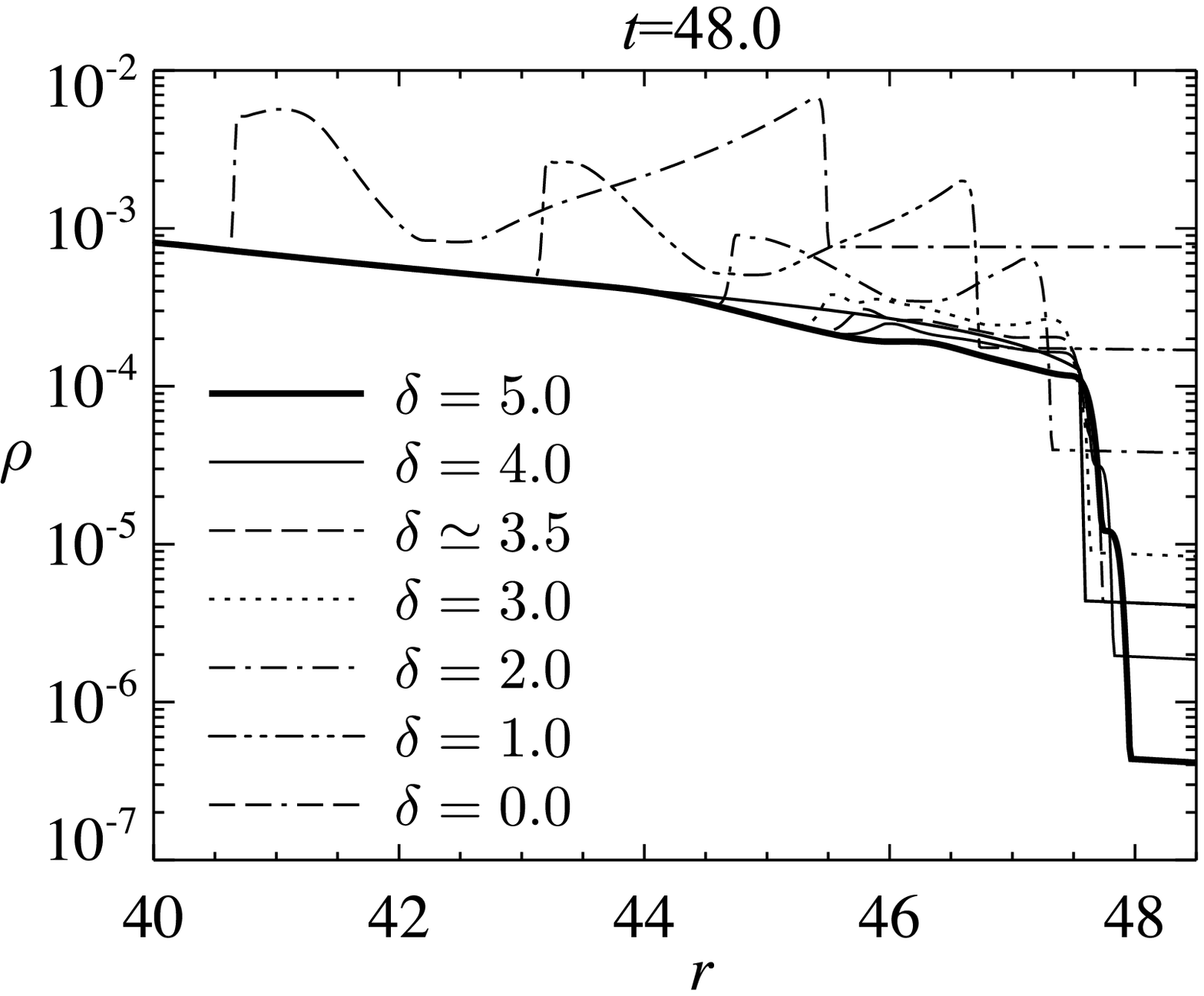}
  \end{minipage}
 \end{tabular}
 \caption{The dependence of the equatorial Lorentz factor (left) and the 
 density profile (right) at $t=48$ on the density parameter of the 
 ambient gas $\delta$. Dashed curves denote the numerical results for 
 initial condition given by the analytical solutions ($\delta\simeq 3.5$). 
 The other curves depict those for $\delta=5, 4, 3, 2, 1, 0$.
 }
 \label{fig:rogm-delta}
\end{figure}

Fig.~\ref{fig:gmevol} shows the time evolution of the Lorentz
factor. Solid curve shows the analytical solution.
We measure the Lorentz factor at a reference radius $r_*=0.995R_s(t)$ to
avoid the effects of the shock flattening. 
Since the simulation is carried out at discrete grid points, we plot the 
range of the Lorentz factors of the mesh points closest to the reference 
radius $r_*=0.995R_s$ (grey contours). The numerical results deviate
from the analytical 
solution when $t<13$ because the flattening of the shock front and 
the emission of the sound waves temporarily increase the velocity 
behind the shock.
The numerical results are, however, 
consistent with the analytical solutions after the sound waves propagate away
($t\ga 18$). Then the Lorentz factor increases with time as $\gamma
\propto t^{1/4}$.

Next we numerically verify the approximation of neglecting the
thermal enthalpy term $h_N$ in the gravitational force $-GM\rho h
\gamma^2/r^2 = -GM\rho (1+h_N)\gamma^2/r^2$ \citep{1986ApJ...309..455M,
2006A&A...447..797M}. When we derive analytical solutions of the
self-similar expansion, we neglect $h_N$ in the gravity.
Although the specific thermal enthalpy $h_N$ becomes larger
than unity behind the strong shock ($p_s/\rho_s\simeq \Gamma_s
\gg 1$, see equations \ref{eq:RHp}-\ref{eq:RHg}), 
the gravitational force itself becomes small compared to the other
forces in the self-similar stage when $R_s \gg r_g$.
To evaluate the contribution of the
gravity, we carried out numerical simulations for three models of the
gravity, i.e., (i) $-GM\rho \gamma^2/r^2$, (ii) $-GM\rho h\gamma^2/r^2$
(iii) without gravity. 

Figure~\ref{fig:gcomp} shows the radial profiles of the Lorentz factor on
the equatorial plane at $t=48$ with different models of the
gravity. Solid curve shows the self-similar solutions, while dashed,
dash-dotted, and dotted curves show the numerical results
for models (i)-(iii), respectively. An inset shows solutions in the different range
of $r$, $1\lid r \lid 25$. Behind the shock, numerical results are almost
independent of the
models of the gravitational force, indicating that the gravitational force
is much smaller than the other forces. As we mentioned in \S~2, the
ratio of the gravity for the thermal enthalpy to the plasma inertia
decreases with radius (see equation \ref{eq:fratio}). The ratio $r_g/R_s$ is
0.01 at the initial state $t=t_0$ and $3\times 10^{-3}$ at the final
state $t=48$ in our simulations. Thus the gravity for the thermal
enthalpy is negligible. This explains why the numerical results are
independent of the gravity models.
The specific thermal enthalpy $h_N$ is larger than unity just behind the
shock, but the gravitational force is much smaller than the other forces
when $r_g \ll R_s$. 

Although the gravity becomes important in the region where $r_g < r \ll
R_s$, the thermal enthalpy is negligible in this region (i.e., $h_N \ll
1$). Thus, we can neglect the contribution of $ h_N $ in the gravity
(see the inset of Fig.~\ref{fig:gcomp}).
 We note that the numerical results for model (iii) 
deviate from the analytical solutions in this region. The plasma is
accelerated in radial direction by the pressure gradient force,
leading to the formation of shocks ($r\simeq 18.5$). 
Since we use analytical solutions for inner boundary conditions at 
$r=R_\mathrm{in} = 1$, the plasma is supplied from
the inner boundary. 
When the plasma is not confined by gravity, the outflowing plasma
forms second shocks at $r \simeq 11$. 

When $R_s$ is close to $r_g$, the contribution of $h_N$ in the gravity is not
negligible, but we have to take into account the general relativistic
effects in such region, so that the characteristic length $r_g$ enters
into the formulations. In such regime, no self-similar solutions can be
obtained. It is out of the scope of this paper to obtain
solutions in this regime.

Next we carried out simulations with different density profiles of
the ambient plasma to study the generality of the analytical solutions
and the effects of the ambient density distribution on the loop dynamics.
We substitute the density profile given in equation (\ref{sol:rho0})
with the power law profile as,
\begin{equation}
 \rho(r) = \rho_0(r=R_{s,0})
  \left(\frac{r}{R_{s,0}}\right)^{-\delta}, \label{eq:ropower}
\end{equation}
where $R_{s,0}=R_s(t=t_0)$.
The analytical solutions correspond to $\delta \simeq 3.5$. We study the
several case ($\delta=0, 1, 2, 3, 4, 5$).
The initial condition is given by the analytical solutions inside the shock.

The dependence of the equatorial Lorentz factor (left) and the density
profile (right) at $t=48$ on the density parameter of the ambient gas
$\delta$ are plotted in Fig.~\ref{fig:rogm-delta}. 
Dashed curve denotes the numerical results for initial conditions given
by the analytical solutions ($\delta \simeq 3.5$). Other curves depict those
for $\delta = 5, 4, 3, 2, 1, 0$.
The peak Lorentz factor decreases as $\delta$ decreases. This is because the
shell becomes massive for a smaller $\delta$ by sweeping up the larger
ambient plasma. The swept up plasma is accumulated behind the shock surface.
The inertia from the excess plasma accumulated behind the shock
decelerates the outflows and creates another discontinuity behind the
shock. The discontinuity can be considered as the reverse shock. The
compression ratio of the reverse shock is larger for the denser ambient
plasma.

\begin{figure}
   \center
 \includegraphics[width=7.5cm]{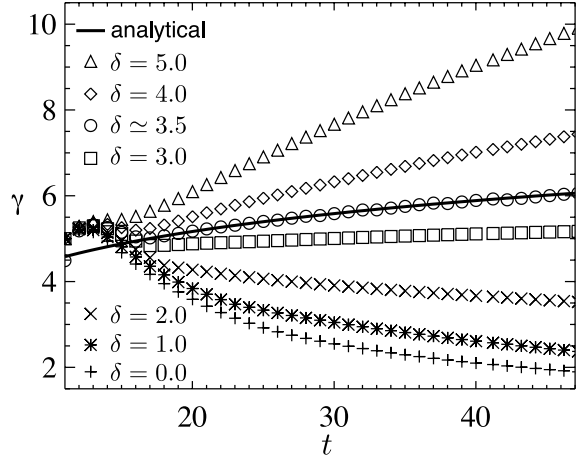}
 \caption{Time evolution of the Lorentz factor at the reference radius 
 $r_*$. Circles depict the numerical results for the initial conditions 
 given by the analytical solution for $\delta\simeq 3.5$. Other symbols 
 show the numerical results for $\delta=5,4,3,2,1,0$.
 The Lorentz factor is evaluated at $r_*=0.995r(\gamma_{\mathrm{max}})$, 
 where $r(\gamma_{\mathrm{max}})$ is the radius where the Lorentz 
 factor is its maximum.}
 \label{fig:gmpeak}
\end{figure}
Time evolution of the equatorial Lorentz factor at the reference
radius $r_*$ are shown in Fig.~\ref{fig:gmpeak}. 
Circles depict the numerical results for the initial conditions given
by the analytical solutions for $\delta \simeq 3.5$. Other symbols show the
numerical results for $\delta = 5, 4, 3, 2, 1, 0$.
The Lorentz factor is evaluated at $r_*=0.995r(\gamma_{\mathrm{max}})$, 
where $r(\gamma_{\mathrm{max}})$ is the radius where the Lorentz 
factor is its maximum.

The peak Lorentz factor increases with time when $t<13$. This increase 
comes from the emission of the sound waves propagating from the shock 
front. When $t>18$, the peak Lorentz factor increases with time when 
$\delta >3$. The critical value of $\delta$ whether the outflow is 
accelerated or not can be evaluated from the mass conservation.
The plasma density of the outflow decreases with radius by
$r^{-3}$ according to the mass conservations (see equation
\ref{eq:rho}). The rest mass energy of the ambient plasma accumulated in
the shell thus increases with time when $\delta<3$. On the other hand,
when $\delta>3$, the outflow is accelerated since the plasma inertia of
the outflow decreases with time.  
Naively, we can understand these processes from equation (\ref{eq:Gs}). 
According to this equation, the flow is accelerated when the ambient
plasma density decreases faster than $r^{-3}$. When $\delta = 5$,
the shock is accelerated and its Lorentz factor is proportional to the
radius ($\simeq$ time).
It indicates that the flow expands freely. The influence of the
ambient plasma is almost negligible. \cite{1993MNRAS.263..861P} derived the
self-similar solution of the free expansion. The shock Lorentz factor
then increases with radius linearly. The shock profile for $\delta=5$ or
a larger $\delta$ is consistent with the solution obtained by
\cite{1993MNRAS.263..861P}.
Inside the shock surface, our solutions are, however, not
identical with their analytical solutions since our solutions
include the magnetic fields and are intrinsically non-spherical.

\section{Summary \& Discussions}\label{summary}
We derived axisymmetric relativistic self-similar solutions of the
magnetic flux rope expansion by assuming the purely radial flow and
ignoring the stellar rotation.
By taking the self-similar variable as $\eta=r/Z(t)$, the arbitrary
function $Z(t)$ has a unique form given in equation (\ref{sol:Z}). The
MHD equations are then solved analytically. 

The solutions obtained in this paper are the extension of our previous work
\citep{2009MNRAS.394..547T} by considering the two
discontinuities, the contact discontinuity and the shock. The contact
discontinuity separates the outflowing plasma and the
ambient plasma. The outgoing waves propagating in the ambient plasma form
shocks. 

Such a self-similar solutions including two discontinuities are derived by
\cite{1984ApJ...281..392L} in non-relativistic MHD. Our solutions are the
extension of their solutions to the relativistic MHD.
For the non-relativistic case, the compression
ratio at the shock is determined by the specific heat ratio. The
specific heat ratio is taken as $\Gamma=4/3$ in non-relativistic MHD
equations. The system is marginally
stable for the inertial flow. The compression ratio is then up to 7 for the
strong shocks. In relativistic plasma, the sound speed
is limited to $\sim 0.58$ for the ideal gas. The differences
between the upstream flow velocity and the downstream wave velocity are
larger for the larger flow velocity. This fact results in forming the
strong shocks and the compression ratio can be larger than 7. The
ambient plasma is abruptly heated by the strong shocks. The hot
plasma is accumulated in the shell between the shocks and the contact
discontinuity. 
Inside the contact discontinuity, the magnetic loops anchored to the
central star are assumed to follow the flux rope solutions obtained by
\cite{1984ApJ...281..392L}. The flux ropes are contained inside the
global magnetic loops. Such magnetic field configuration can be expected
for the SGR flares after the magnetic energy is dissipated
\citep{2006MNRAS.367.1594L}.

We also carried out numerical simulations of two dimensional
relativistic MHD by using the self-similar solutions as the initial and
inner boundary conditions. Since analytical solutions are
obtained in this paper, we can apply them to check the accuracy of 
multi-dimensional relativistic MHD codes. 
Many previous authors reported that the relativistic MHD code is
verified by using the one dimensional shock tube problems. There are
only a few standard multi-dimensional problems, such as the blast wave
problem or the rotor problem. However, no exact solutions are known for
these problems. 
We have shown that the self-similar solutions can be
applied to check
the accuracy of the relativistic MHD codes.

Numerical calculations show that the shock velocity strongly depends on
the ambient plasma density. 
When the density profile is steeper than $\propto r^{-3}$, the shock
Lorentz factor increases with radius. On the other hand, it decreases
for profiles shallower than $r^{-3}$. 
Such a behavior is expected from the consideration of the mass
conservation \citep{1979ApJ...233..831S}. 
The density inside the shocks approximately decreases with radius as
$\propto r^{-3}$. When the decrease in the ambient plasma density
is steeper than the shocked plasma density, the outflows
can be accelerated.
We can understand this from the energy conservation given in equation
(\ref{eq:Gs}). The time dependence of the shock Lorentz factor is
related to the ambient plasma density. When $R_s\simeq t$ and
the ambient plasma density is represented by the power law on $r$
($\rho\propto r^{-\delta}$), the shock Lorentz factor is expressed as 
$\Gamma_s \propto r^{(3-\delta)/2}$. Numerical results agree with this
relation. Especially when $\delta \gid 5$, the shock Lorentz factor
linearly increases with radius.

Finally let us apply our results to the magnetar flares
\citep{2006csxs.book..547W, 2008A&ARv..15..225M}. 
In the discussion below, 
we concentrate on the extraordinarily energetic outbursts (giant flares)
observed in SGRs.
Although AXPs as well as SGRs would be magnetars, 
the giant flares have not been detected from the AXPs.
The reason would be that the giant flares are very rare events (1 per
a few decades in the whole sky).

The shock Lorentz factor $\Gamma_s$ is estimated from 
the mass ejected by the flare $M_\mathrm{eje}\sim 4\pi \rho_0
R_s^3/3$, and equation (\ref{sol:nonE}) as
\begin{equation}
 \Gamma_s \simeq 24
  \left(\frac{\mathcal{E}'}{10^{46}~\mathrm{erg}}\right)^{\frac{1}{2}}
  \left(\frac{M_\mathrm{eje}}{10^{22}~\mathrm{g}}\right)^{-\frac{1}{2}}.\label{eq:Gsn}
\end{equation}
The expansion speed of the magnetic loop is relativistic when $M_\mathrm{eje} <
10^{22}~\mathrm{g}$ \citep[see, also][]{2006MNRAS.367.1594L}.
The mass density behind the forward shock is estimated from
equations (\ref{eq:RHr}), (\ref{eq:RHg}) and (\ref{eq:Gsn}) as 
\begin{equation}
 \rho_s = 1.6 \times 10^{-4} ~\mathrm{g~cm^{-3}}~
  \left(\frac{\mathcal{E}'}{10^{46}~\mathrm{erg}}\right)^{\frac{1}{2}}
  \left(\frac{M_\mathrm{eje}}{10^{22}~\mathrm{g}}\right)^{\frac{1}{2}}
  \left(\frac{R_s}{10^9~\mathrm{cm}}\right)^{-3}.\label{eq:rosn}
\end{equation}
The mass density in the ambient plasma $\rho_0=\rho_s/(2^{3/2}\Gamma_s)$
estimated from equations
(\ref{eq:RHr}), (\ref{eq:RHg}), (\ref{eq:Gsn}) and (\ref{eq:rosn}), is much
larger than the Goldreich-Julian density. 
Such dense coronal pair plasmas would be created by the pair production
when the magnetically trapped fireball is formed on the surface
of the magnetar \citep{2007ApJ...657..967B}. 

The temperature
of the shocked gas is evaluated by using equation (\ref{eq:RHp}) as
\begin{equation}
 T_s\simeq 2.8~\mathrm{MeV}~
  \left(\frac{\mathcal{E}'}{10^{46}~\mathrm{erg}}\right)^{\frac{1}{2}}
  \left(\frac{M}{10^{22}~\mathrm{g}}\right)^{-\frac{1}{2}}.\label{eq:Tsn}
\end{equation}
Here we assume the pair plasma. 
Although the radiation spectrum of the initial spike in the giant flare
is not well determined because of its short duration, the
typical temperature indicated by the spectrum is a few
$100~\mathrm{keV}$ \citep{2005Natur.434.1098H}, which is
lower than $T_s$ estimated from our analytical solutions. We
have to note that the temperature at the photosphere $T_\mathrm{ps}$
should be smaller than $T_s$
because the temperature decreases with decreasing the radius behind
the shock (see, Fig.~\ref{fig:hydro-r} and Fig.~\ref{fig:rogm}). 
The radius of the photosphere $R_\mathrm{ps}$ can be determined by the
condition that the optical depth of the expanding magnetic loops
$\tau_\mathrm{ps}$ satisfies
\begin{equation}
 \tau_\mathrm{ps} = 
  \int_{R_\mathrm{ps}}^{R_s}~
  \rho \tilde \kappa \gamma (1-v\cos\theta)dr=1,\label{eq:tau}
\end{equation}
where $\theta$ is the angle between
the velocity vector and the direction of the photon propagation
\citep{1991ApJ...369..175A}, and $\tilde \kappa =
\sqrt{\kappa_\mathrm{ff}(\kappa_\mathrm{es}+\kappa_\mathrm{ff})}$ is the
effective opacity. Here $\kappa_\mathrm{es}$ and $\kappa_\mathrm{ff}$
are the opacity for the electron scattering and the free-free
absorption, respectively. 
We assume that the ambient plasma ($r>R_s$) is optically thin. 
The electron scattering is the dominant source for the opacity inside
the shocks. This is because the plasma temperature is increased by
the shock heating, so that the free-free opacity ($\rho T^{-3.5}$) is
much smaller than that of the electron scattering.

We numerically
integrate equation (\ref{eq:tau}) assuming that $\theta=0$, the expansion energy
$\mathcal{E}'=10^{46}~\mathrm{erg}$, and the mass of the central star
$M=1M_\odot$, where $M_\odot$ is the solar mass.
The temperature at the photosphere $T_\mathrm{ps}$ calculated at
$R=R_\mathrm{ps}$ from the analytical solutions (equations \ref{eq:Zt}
\ref{eq:p}, \ref{eq:rho}, \ref{sol:Ps} and \ref{sol:Ds}) can
roughly be fitted by
\begin{equation}
 T_\mathrm{ps}\simeq 10 \left(\frac{\Gamma_s}{10}\right)^{-1.4}
  \left(\frac{R_s}{10^9~\mathrm{cm}}\right)^{-1.3}~\mathrm{keV},\label{eq:Tpsn}
\end{equation}
when $10\la \Gamma_s \la 100$ and $10^7~\mathrm{cm}\la R_s \la
10^{10} ~\mathrm{cm}$. 
The result is almost independent of the radius of the contact
discontinuity (i.e., $R_c/R_s$). 
The temperature in the observer's frame is $T_\mathrm{obs}=\gamma
T_\mathrm{ps}\simeq \Gamma_s T_\mathrm{ps}\simeq 100\mathrm{keV}$ when
$\Gamma_s\simeq 10$. Here we assume $\gamma\simeq \Gamma_s$ since the
photosphere is very close to the shock surface.
This result is consistent with the
observational results when $\Gamma_s\simeq 10$ and $R_s\simeq
10^{9}~\mathrm{cm}$.

Such hot, relativistically expanding magnetic loops are expected to be
formed with the help of the magnetic reconnections.
According to the scenario by \cite{2006MNRAS.367.1594L}, the magnetic
reconnections inside the magnetic loops are responsible for the initial
spike of the flares \citep[see, also][]{2010MNRAS.407.1926G}. 
Subsequently, the plasma is heated up by the shocks produced by the
magnetic reconnection.
By applying our models to the giant flares, 
the time evolution of the luminosity from expanding plasma is expressed
as $L\propto T_\mathrm{obs}^4 R_\mathrm{ps}^2\propto \Gamma_s^{-1.6}
R_s^{-3.2}$ from $T_\mathrm{obs} \sim \Gamma_s T_\mathrm{ps}$ and equation
(\ref{eq:Tpsn}). Here we assumed $R_\mathrm{ps} \simeq R_s$.
Since $\Gamma_s \propto R_s^{(\delta-3)/2}$ from equations (\ref{eq:Gs})
and (\ref{eq:ropower}), we obtain $L \propto t^{-0.8(1+\delta)}$
because $R_s \propto t$ from equation (\ref{sol:appRs}). 
\cite{2005Natur.434.1110T} reported that the
observed photon counts decreased exponentially with time after the
initial spike. However, the photon counts at the earlier stage, whose
decay time is very short ($\la 100~\mathrm{ms}$), is not inconsistent
with the power-law decay.

In addition to the initial spikes, a hump is observed a few
hundred seconds after the initial spike \citep{2005Natur.434.1110T}. It is
considered that the energy is re-injected from the central star. 
We now consider another possibility for this hump.
Before the magnetic energy release, the toroidal magnetic energy can be
comparable to that of the poloidal magnetic fields inside the magnetic
loops. When the magnetic reconnection takes place inside the magnetic
loops, the magnetic energy of the poloidal magnetic fields is converted
to the plasma energies, generating the twisted flux ropes. The
toroidal magnetic field energy of the flux ropes does not dissipate in
this process. 
When the twisted flux ropes cross the Alfv\'en radius ($\simeq$
light cylinder), the rest magnetic energy will be dissipated by
interaction with the ambient global magnetic fields. 
The time scale that a flux rope crosses the light cylinder and releases
the magnetic energy is about $\sim 1~\mathrm{sec}$, which is consistent
with the observations. The
energy dissipation at the light cylinder can be responsible for the humps.
We need further study to verify these processes.

\section*{Acknowledgments}
We are grateful to an anonymous referee for improving our manuscript.
We thank Hiroaki Isobe, Jin Matsumoto, Kazunari Shibata, and Youhei
Masada for useful discussions. 
Numerical computations were carried out on Cray XT4 at Center for
Computational Astrophysics, CfCA, of National Astronomical Observatory
of Japan and on Fujitsu FX-1 at JAXA Supercomputer System (JSS) of Japan
Aerospace Exploration Agency (JAXA).
This work was supported by the
Grants-in-Aid for Scientific Research of Ministry of Education, Culture,
Sports, Science, and Technology (RM:20340040).


\label{lastpage} 


\begin{thebibliography}{37}
\expandafter\ifx\csname natexlab\endcsname\relax\def\natexlab#1{#1}\fi

\bibitem[{{Abramowicz} {et~al.}(1991){Abramowicz}, {Novikov}, \&
  {Paczynski}}]{1991ApJ...369..175A}
{Abramowicz} M.~A., {Novikov} I.~D., {Paczynski} B., 1991, \apj, 369, 175

\bibitem[{{Asano}(2007)}]{2007Chiba...1..1}
{Asano} E., 2007, PhD. thesis, Chiba Univ.

\bibitem[{{Beloborodov} \& {Thompson}(2007)}]{2007ApJ...657..967B}
{Beloborodov} A.~M., {Thompson} C., 2007, \apj, 657, 967

\bibitem[{{Blandford} \& {McKee}(1976)}]{1976PhFl...19.1130B}
{Blandford} R.~D., {McKee} C.~F., 1976, Physics of Fluids, 19, 1130

\bibitem[{{Burrows}(1987)}]{1987ApJ...318L..57B}
{Burrows} A., 1987, \apjl, 318, L57

\bibitem[{{Duncan} \& {Thompson}(1992)}]{1992ApJ...392L...9D}
{Duncan} R.~C., {Thompson} C., 1992, \apjl, 392, L9

\bibitem[{{Gill} \& {Heyl}(2010)}]{2010MNRAS.407.1926G}
{Gill} R., {Heyl} J.~S., 2010, \mnras, 407, 1926

\bibitem[{{Gourgouliatos} \& {Lynden-Bell}(2008)}]{2008MNRAS.391..268G}
{Gourgouliatos} K.~N., {Lynden-Bell} D., 2008, \mnras, 391, 268

\bibitem[{{Gourgouliatos} \& {Vlahakis}(2010)}]{2010arXiv1001.4209G}
{Gourgouliatos} K.~N., {Vlahakis} N., 2010, ArXiv e-prints

\bibitem[{{Harten} {et~al.}(1983){Harten}, {Lax}, \& {van
  Leer}}]{1983siamRev...25..35..61}
{Harten} A., {Lax} P.~D., {van Leer} B., 1983, SIAM Rev., 25, 35

\bibitem[{{Hurley} {et~al.}(2005){Hurley}, {Boggs}, {Smith}, {Duncan}, {Lin},
  {Zoglauer}, {Krucker}, {Hurford}, {Hudson}, {Wigger}, {Hajdas}, {Thompson},
  {Mitrofanov}, {Sanin}, {Boynton}, {Fellows}, {von Kienlin}, {Lichti}, {Rau},
  \& {Cline}}]{2005Natur.434.1098H}
{Hurley} K., {Boggs} S.~E., {Smith} D.~M., {Duncan} R.~C., {Lin} R., {Zoglauer}
  A., {Krucker} S., {Hurford} G., {Hudson} H., {Wigger} C., {Hajdas} W.,
  {Thompson} C., {Mitrofanov} I., {Sanin} A., {Boynton} W., {Fellows} C., {von
  Kienlin} A., {Lichti} G., {Rau} A., {Cline} T., 2005, \nat, 434, 1098

\bibitem[{{Keil} {et~al.}(1996){Keil}, {Janka}, \&
  {Mueller}}]{1996ApJ...473L.111K}
{Keil} W., {Janka} H.-T., {Mueller} E., 1996, \apjl, 473, L111+

\bibitem[{{Kennel} \& {Coroniti}(1984)}]{1984ApJ...283..710K}
{Kennel} C.~F., {Coroniti} F.~V., 1984, \apj, 283, 710

\bibitem[{{Low}(1982{\natexlab{a}})}]{1982ApJ...254..796L}
{Low} B.~C., 1982{\natexlab{a}}, \apj, 254, 796

\bibitem[{{Low}(1982{\natexlab{b}})}]{1982ApJ...261..351L}
---, 1982{\natexlab{b}}, \apj, 261, 351

\bibitem[{{Low}(1984{\natexlab{a}})}]{1984ApJ...281..392L}
---, 1984{\natexlab{a}}, \apj, 281, 392

\bibitem[{{Low}(1984{\natexlab{b}})}]{1984ApJ...281..381L}
---, 1984{\natexlab{b}}, \apj, 281, 381

\bibitem[{{Lyutikov}(2002)}]{2002PhFl...14..963L}
{Lyutikov} M., 2002, Physics of Fluids, 14, 963

\bibitem[{{Lyutikov}(2006)}]{2006MNRAS.367.1594L}
---, 2006, \mnras, 367, 1594

\bibitem[{{Lyutikov} \& {Blandford}(2003)}]{2003astro.ph.12347L}
{Lyutikov} M., {Blandford} R., 2003, astro-ph/0312347

\bibitem[{{Meliani} {et~al.}(2006){Meliani}, {Sauty}, {Vlahakis}, {Tsinganos},
  \& {Trussoni}}]{2006A&A...447..797M}
{Meliani} Z., {Sauty} C., {Vlahakis} N., {Tsinganos} K., {Trussoni} E., 2006,
  \aap, 447, 797

\bibitem[{{Mereghetti}(2008)}]{2008A&ARv..15..225M}
{Mereghetti} S., 2008, \aapr, 15, 225

\bibitem[{{Mignone} \& {Bodo}(2006)}]{2006MNRAS.368.1040M}
{Mignone} A., {Bodo} G., 2006, \mnras, 368, 1040

\bibitem[{{Mobarry} \& {Lovelace}(1986)}]{1986ApJ...309..455M}
{Mobarry} C.~M., {Lovelace} R.~V.~E., 1986, \apj, 309, 455

\bibitem[{{Piran} {et~al.}(1993){Piran}, {Shemi}, \&
  {Narayan}}]{1993MNRAS.263..861P}
{Piran} T., {Shemi} A., {Narayan} R., 1993, \mnras, 263, 861

\bibitem[{{Prendergast}(2005)}]{2005MNRAS.359..725P}
{Prendergast} K.~H., 2005, \mnras, 359, 725

\bibitem[{{Sari}(2006)}]{2006PhFl...18b7106S}
{Sari} R., 2006, Physics of Fluids, 18, 027106

\bibitem[{{Shapiro}(1979)}]{1979ApJ...233..831S}
{Shapiro} P.~R., 1979, \apj, 233, 831

\bibitem[{{Spitkovsky}(2005)}]{2005KITP...CONF..HP}
{Spitkovsky} A., 2005, KITP conference Physics of Magnetized rotators: Force
  Free Electrodynamics Simulations http://online.kitp.ucsb.edu/online/

\bibitem[{{Stone} {et~al.}(1992){Stone}, {Hawley}, {Evans}, \&
  {Norman}}]{1992ApJ...388..415S}
{Stone} J.~M., {Hawley} J.~F., {Evans} C.~R., {Norman} M.~L., 1992, \apj, 388,
  415

\bibitem[{{Stone} \& {Norman}(1992)}]{1992ApJS...80..753S}
{Stone} J.~M., {Norman} M.~L., 1992, \apjs, 80, 753

\bibitem[{{Takahashi} {et~al.}(2009){Takahashi}, {Asano}, \&
  {Matsumoto}}]{2009MNRAS.394..547T}
{Takahashi} H.~R., {Asano} E., {Matsumoto} R., 2009, \mnras, 394, 547

\bibitem[{{Terasawa} {et~al.}(2005){Terasawa}, {Tanaka}, {Takei}, {Kawai},
  {Yoshida}, {Nomoto}, {Yoshikawa}, {Saito}, {Kasaba}, {Takashima}, {Mukai},
  {Noda}, {Murakami}, {Watanabe}, {Muraki}, {Yokoyama}, \&
  {Hoshino}}]{2005Natur.434.1110T}
{Terasawa} T., {Tanaka} Y.~T., {Takei} Y., {Kawai} N., {Yoshida} A., {Nomoto}
  K., {Yoshikawa} I., {Saito} Y., {Kasaba} Y., {Takashima} T., {Mukai} T.,
  {Noda} H., {Murakami} T., {Watanabe} K., {Muraki} Y., {Yokoyama} T.,
  {Hoshino} M., 2005, \nat, 434, 1110

\bibitem[{{Uchida}(1997)}]{1997PhRvE..56.2181U}
{Uchida} T., 1997, \pre, 56, 2181

\bibitem[{{van Leer}(1977)}]{1977JCoPh..23..263V}
{van Leer} B., 1977, Journal of Computational Physics, 23, 263

\bibitem[{{Woods} {et~al.}(2001){Woods}, {Kouveliotou}, {G{\"o}{\u g}{\"u}{\c
  s}}, {Finger}, {Swank}, {Smith}, {Hurley}, \&
  {Thompson}}]{2001ApJ...552..748W}
{Woods} P.~M., {Kouveliotou} C., {G{\"o}{\u g}{\"u}{\c s}} E., {Finger} M.~H.,
  {Swank} J., {Smith} D.~A., {Hurley} K., {Thompson} C., 2001, \apj, 552, 748

\bibitem[{{Woods} \& {Thompson}(2006)}]{2006csxs.book..547W}
{Woods} P.~M., {Thompson} C., 2006, in Lewin W., van der Klis M., eds,
  Cambridge Astrophys. Ser. Vol. 39, Compact stellar X-ray sources, Cambridge
  Univ. Press, Cambridge, p. 547

\end{thebibliography}
\end{document}